\documentclass[preprint, 12pt, numbered]{elsarticle}

\usepackage{color}
\usepackage{a4wide}
\usepackage{amssymb}
\usepackage{amsmath}
\usepackage{graphicx}
\usepackage{mathrsfs}
\usepackage{caption}
\usepackage{subcaption}
\usepackage{textcomp}
\usepackage{multirow}
\usepackage{array}
\usepackage{enumitem}
\usepackage{hyperref}
\hypersetup{
  colorlinks,
  citecolor=blue,
  filecolor=green,
  linkcolor=cyan,
  urlcolor=purple
}
\begin{document}
\begin{frontmatter}
%\date{}
\title{Modified symmetry technique for mitigation of flow leak near corners for compressible inviscid fluid flow}
\author[add1]{Vinnakota Mythreya\corref{cor1}\fnref{fn1}}
\cortext[cor1]{Corresponding Author}
\ead{myth.vinna007@gmail.com}
\fntext[fn1]{Doctoral Candidate}

\author[add1]{ M. Ramakrishna\fnref{fn2}}
\ead{krishna@ae.iitm.ac.in}
\fntext[fn2]{Professor}

\address[add1]{Department of Aerospace Engineering, IIT Madras}

%\begin{document}
%\maketitle
\begin{abstract}
  Using the standard symmetry technique for applying boundary conditions for free slip and flat walls with corners will lead to flow leak through the wall near corners 
  (violation of no penetration condition) 
  and a corresponding error in prediction of pressure. Also, prescribing a state at the corner as a boundary condition is not possible.
  In this paper, a method for tackling the `corner point state' problem is given and modifications to the standard symmetry technique are proposed to mitigate flow leak near
  the
  corner. Using this modified symmetry technique, numerical solutions to the Euler equations for flows over forward facing and backward facing step are computed employing the
  Shu-Osher conservative finite difference
  scheme with WENO-NP3 reconstruction (with a formal order of accuracy in space of 3), Lax-Freidrichs flux splitting, and TVD-RK3 time discretisation.
  It is shown that using this
  modified symmetry technique leads to mitigation of flow leak near the corner and a better prediction of shock structure even on coarse meshes. 
  \end{abstract}
\end{frontmatter}
\section{Introduction} \label{Sec:Introduction}
Problems with corners arise while computing numerical solutions to several differential equations like the heat equation, the incompressible Navier-Stokes equations and
the compressible Euler equations. Applying boundary conditions at and near a corner is often an issue and it was addressed in many papers, some of which are:
\begin{itemize}
  \item J. Crank et al \cite{crank1978} listed various strategies employed in tackling corners for linear elliptic and parabolic equations,
  \item H. Holstein. Jr \cite{holsteinJr1981}, G. A. Ache \cite{acheUnivWisconson1987}, addressed the re-entrant corner issue for incompressible viscous fluid flow equations,
  \item Woodward et al \cite{woodward1984} pointed out that the expansion corner of a free slip flat wall in the Mach 3.0 flow over a forward facing step is a singular point
    and they proposed a way for tackling problems that arise due to it,
  \item A. Verhoff \cite{verhoff2004} used analytic methods to address the problem of corners for compressible inviscid fluid flows, which were modelled using an approximation
    of the Euler equations.
\end{itemize}
In this paper, we revisit solutions to the compressible Euler equations. We describe problems with applying boundary conditions associated with free slip flat walls having
corners that are not known to be stagnation points. We propose ways to address these problems.
%***Improve the previous sentence and try to get `free slip flat walls having corners' to the end of the sentence***

Free slip condition corresponds to `no resistance to the flow at the boundary'. Solid and rigid wall boundary conditions are equivalent to a no penetration condition at
the boundary.
To enforce these conditions in computations, the wall is often assumed to be a surface of symmetry. Henceforth, the technique of enforcing the wall boundary to be a
surface of symmetry will be referred to as the `standard symmetry technique' or `SST'.

One issue with using SST is that it is accurate only for flat walls. Using SST leads to a zero pressure gradient
normal to the wall ($\partial p/\partial n$), which is not always the case on a curved wall as pointed out by Moretti \cite{moretti1969}. For example, in the two-dimensional 
flow over a wall with a radius of curvature $`R$', $\partial p/\partial n = -(\rho V_{\tau}^2)/R$ (where $V_{\tau}$ is the component of the velocity $\vec{V}$ that is 
tangential to the wall)
is non-zero and therefore using SST will lead to inaccuracy in the numerical
solution. The problem of applying an appropriate pressure gradient on the wall was addressed using different techniques, one amongst which is the `curvature corrected symmetry
technique', introduced by  Dadone \cite{dadone1993, dadone1994}.

Unfortunately, these techniques do not address the case of zero radius of curvature, namely a wall with a corner.
A problem with the corner was addressed by Verhoff \cite{verhoff2004, verhoff2006}, using analytic methods for solving approximations to steady, two-dimensional,
compressible Euler equations, written in streamline coordinates.
Approximate solutions for various problems including that of subsonic flow over a ramp \cite{verhoff2004},
which has both compression and expansion corners, were reported.
The first approximate solution obtained for the flow over a ramp predicted infinite momentum density at expansion corner and zero momentum
density at compression corner.
To tackle the singularity and obtain density, pressure, and magnitude of momentum, coordinate straining was used, which the author pointed out was `arbitrary to some extent'.
Applying boundary conditions for such walls with corners is problematic for two reasons. 

\begin{figure}[!htbp]
  \centering
  \begin{subfigure}{.30\textwidth}
    \centering
  \includegraphics[width=.9\textwidth]{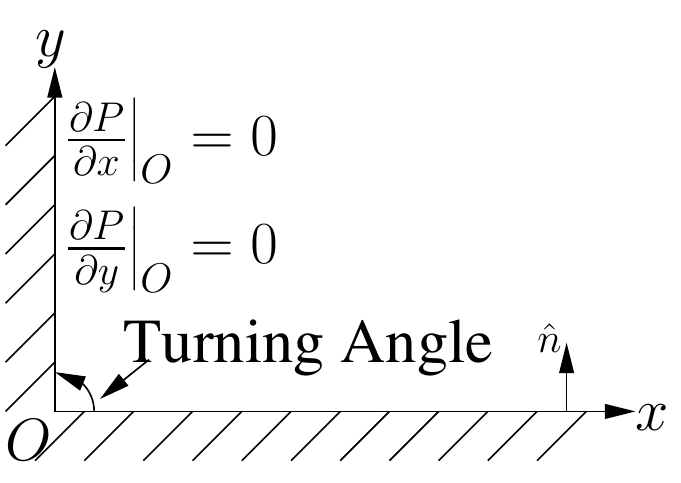}
  \caption{Compression corner}
  \label{fig:compressionCorner}
\end{subfigure}%
\begin{subfigure}{.5\textwidth}
  \centering
  \includegraphics[width=.9\textwidth]{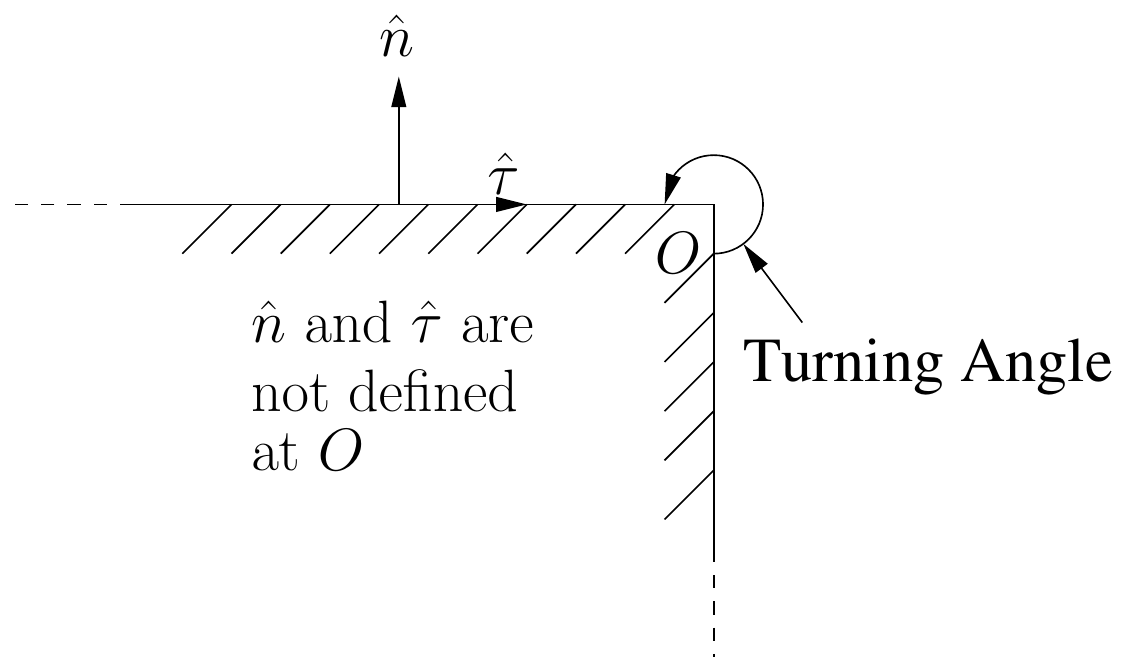}
  \caption{Expansion corner}
  \label{fig:expansionCorner}
\end{subfigure}
\caption{Types of corners. $\hat{n}$, the normal to the boundary, is not defined at $O$, in both the cases.}
\label{fig:typesOfCorners}
\end{figure}

First, the normal, $\hat{n}$, and the tangent, $\hat{\tau}$, (see figure~\ref{fig:typesOfCorners}) at the corner are not defined. Consequently, the normal and tangential
components of velocity, ($V_n$, $V_{\tau}$) and ($\partial p/\partial n$) can not be determined. %So what? This is true if we draw a corner at any point in the flow field
Setting $\vec{V}=\vec{0}$ is not a solution, as the corner is not known to be a stagnation point.
Simultaneous application of free slip and no penetration conditions does not seem possible at the corner. Further, in the pressure gradient equation
($\partial p/\partial n = -(\rho V_{\tau}^2)/R$) the radius of
curvature, $R$, is zero. Hence using these symmetry techniques, it is not possible to determine the state at the corner.%determine or define? Is it possible to determine
%state at corner using another technique? Is it possible for the corner to have a well defined state?

%A procedure for calculating the momentum density, density and pressure at the compression and expansion corners was proposed. Some ideas from Verhoff's
%paper \cite{verhoff2004} will be used and will be described later

Second, using SST near a corner is inaccurate simply because, the wall, as will be shown in this paper, is not a surface of symmetry in the neighbourhood of the corner. A free slip wall boundary
condition should preserve the condition of no penetration on the wall. In particular, the pressure on the wall should be such that it does not allow for flow penetration. It
will be shown later that using SST will lead to flow leak through the wall near the corner. This implies that the pressure on the
wall, in the neighborhood of the corner, obtained using SST, is erroneous.

In conclusion, there are two problems for free slip walls with corners, namely, tackling the `corner point state' problem and ensuring no penetration on the wall in the neighborhood of the
corner. These problems will be addressed in this paper.

Based on turning angle of the wall, we classify two-dimensional corners into compression and expansion corners as shown in figure~\ref{fig:typesOfCorners}. This study is
restricted to Cartesian meshes and finite difference schemes. Consequently, turning angles are restricted to
$\pi/2$ and $3 \pi/2$ for compression and expansion corners, respectively. 
In case of a compression
corner with turning angle of $\pi/2$, the velocity and the pressure gradient at the corner are zero. Therefore, there will be no error if SST is used near the corner.
This leaves the problem of applying boundary conditions for expansion corners.

Two of the widely used test problems having walls with such expansion corners are, flow over a backward facing step and flow over a
forward facing step. Numerical solution of flow over a backward facing step was published in a paper by Schmidt and Jameson \cite{schmidt1982}.
Numerical solution of the Mach 3.0 flow over forward facing step (20\% step height) was published in various papers, starting with Ashley F. Emery \cite{emery1968} and later
%in \cite{leer1979, woodward1984, jiang1996, cockburn1998, balsara2009, fan2014}. 
%in \cite{leer1979, woodward1984, jiang1996, cockburn1998, abgrall2006, capdeville2011, sun2016}. 
in \cite{leer1979, woodward1984, jiang1996, cockburn1998}. 
Steady state solution of that problem was published by A. F. Emery and
Bram Van Leer \cite{emery1968, leer1979}, amongst others. In many of these papers, boundary conditions for walls with corners were applied using SST.
For ease of demonstration of problems with using SST near corners, such as flow leak, we use the problem of Mach 4.0 flow over a forward facing step with 20\% step height as
this can be solved on a smaller domain, requiring less computational effort. 

In this paper, we show that using SST for applying boundary conditions on walls near corners allows the violation of the no-penetration condition at and near the corner.
We propose a method to tackle the `corner point state' problem and modifications to the standard symmetry technique to
ensure the no penetration condition is not violated near the corner. A solver employing the Shu-Osher conservative finite difference scheme with WENO-NP3 reconstruction,
Lax-Freidrichs flux splitting and TVD-RK3 time discretisation is developed. This solver is used to compute supersonic flows over forward facing and backward facing step,
using various boundary conditions near the corner and a comparison of the results obtained is presented here.

The rest of this paper is organized as follows. In section~\ref{Sec:NumMethod}, the numerical method employed along with the standard symmetry technique is described. Using
this scheme, the numerical solution of two test problems were computed for verification and validation, and the results obtained are presented. In section
\ref{Sec:DemonstrateLeak}, the problem definition for flow over a forward facing step is given and mass leak is demonstrated.
In section~\ref{Sec:analOfSymmConds}, the cause for the flow leak near expansion corners and ways to mitigate it are discussed. In section~\ref{Sec:MST}, modifications to
the standard symmetry technique are proposed to reduce mass leak
near corners, and in section~\ref{Sec:BCsTesting}, the modified symmetry technique is employed to obtain numerical solutions of Euler equations for flows over forward facing
and backward facing step.

We begin with a brief description of the numerical method, the standard symmetry technique and their implementation.

\section{ Numerical method with the standard symmetry technique }\label{Sec:NumMethod}
In this section, the Shu-Osher conservative finite difference scheme with Weighted Essentially Non-oscillatory (WENO) reconstruction, Lax-Freidrichs flux splitting, and
Total Variation Diminishing-Three stage Runge Kutta (TVD-RK3) time discretisation is described.
This numerical method is used for solving all the problems presented in this paper. The standard symmetry technique used for simulating free slip flat walls is also
described. We start with the description of the Shu-Osher conservative finite difference scheme.

\subsection{Shu-Osher Conservative finite difference scheme}\label{Sec:shuOsherConsFinDiffSch}
Consider a hyperbolic conservation law of the form
\begin{equation}
  \label{eq:hypConsLaw}
  \frac{\partial }{\partial t}Q(x,t) + \frac{\partial}{\partial x}E(Q(x, t)) = 0
\end{equation}
Let the computational domain consist of grid points uniformly spaced in the physical domain, with grid point spacing equal to $\Delta x$. A function $h(x, t)$ is defined such that the sliding average
of $h(x, t)$ over a
length $\Delta x$ is equal to $E(x, t)$, that is,
\begin{equation}
  \label{eq:ImpFn}
  \frac{1}{\Delta x}\int\limits_{-\frac{\Delta x}{2}}^{\frac{\Delta x}{2}}h(x+y, t)dy = E(x, t)
\end{equation}
Taking a partial derivative of equation~(\ref{eq:ImpFn}) with $x$, we get
\begin{equation}
  \label{eq:derivInTermsOfImpFn}
  \frac{\partial E}{\partial x}\bigg|_{x=x_o} = \frac{h(x_o+\frac{\Delta x}{2}, t) - h(x_o-\frac{\Delta x}{2}, t)}{\Delta x}
\end{equation}
We refer to Barry Merriman \cite{merriman2003} for detailed explanation and analysis of the Shu-Osher conservative finite difference scheme.

Using the method of lines and equations~(\ref{eq:hypConsLaw}), and (\ref{eq:derivInTermsOfImpFn}), a semi-discrete form of equation~(\ref{eq:hypConsLaw}) is obtained at
$x=x_o, t=t_o$,
which is
\begin{equation}
  \label{eq:discreteHypConsLaw}
  \frac{\partial Q}{\partial t}\bigg|_{x=x_0, t=t_o} + \frac{h(x_o+\frac{\Delta x}{2}, t_o) - h(x_o-\frac{\Delta x}{2}, t_o)}{\Delta x} = 0
\end{equation}

The global Lax-Freidrichs flux splitting is described below.
\subsection{Upwinding and Flux-Splitting}
To account for propagation along the characteristic directions,
%in the numerical scheme,
upwind biasing of spatial derivatives is needed. This can be achieved by using flux splitting and appropriate biasing of the split fluxes.
For flux splitting, we use the global Lax-Freidrichs flux splitting which is given below:
\begin{equation}
  \label{eq:FSplitting}
  E^{\pm} = \frac{1}{2}\left(E(Q) \pm \alpha Q\right),
\end{equation}
where,
\begin{equation}
  \label{eq:DissipAlpha}
  \alpha = \max_{Q}(|\vec{V}| + a),
\end{equation}
where $a$ is the speed of sound and the maximum is taken over all the grid points in the computational domain. The semi-discrete form of the hyperbolic conservation law incorporating flux
splitting becomes 
\begin{equation}
  \label{eq:discreteHypConsLawFS}
  \frac{\partial Q}{\partial t}\bigg|_{x=x_0, t=t_o} + \frac{h^{+}(x_o+\frac{\Delta x}{2}, t_o) - h^{+}(x_o-\frac{\Delta x}{2}, t_o)}{\Delta x} + \frac{h^{-}(x_o+\frac{\Delta x}{2}, t_o) - h^{-}(x_o-\frac{\Delta x}{2}, t_o)}{\Delta x}= 0,
\end{equation}
where
\begin{equation}
  \label{eq:ImpFnFS}
  \frac{1}{\Delta x}\int\limits_{-\frac{\Delta x}{2}}^{\frac{\Delta x}{2}}h^{\pm}(x+y, t)dy = E^{\pm}(x, t).
\end{equation}
The WENO-NP3 reconstruction procedure is used to obtain approximations to $h^{+}$ and $h^{-}$ using left and right biased stencils, respectively. It is described next.
\subsection{WENO-NP3 reconstruction procedure}\label{Sec:WENOJS3}

WENO-NP3 is one of the family of WENO reconstruction procedures. WENO reconstruction was introduced by Liu, Osher and Chan in 1994 \cite{liu1994}.
Jiang et al gave a framework to build high order WENO schemes \cite{jiang1996}. Changes to these schemes were proposed \cite{borges2008, castro2011, feng2014} to avoid 
loss of accuracy near critical points. One such scheme is the third order WENO-NP3 proposed by Wu et al \cite{wu2016}, which maintains third order accuracy at critical points
also. This will be briefly described below.

%We choose Shu-Osher conservative finite difference scheme with global Lax-Freidrichs flux splitting,  WENO-NP3 reconstruction and TVD-RK3 time discretisation as the numerical 
%scheme to obtain all the reported numerical solutions in this paper. The WENO-NP3 reconstruction is briefly described below.

\begin{figure}[!htbp]
\begin{center}
  \includegraphics[width=0.9\textwidth]{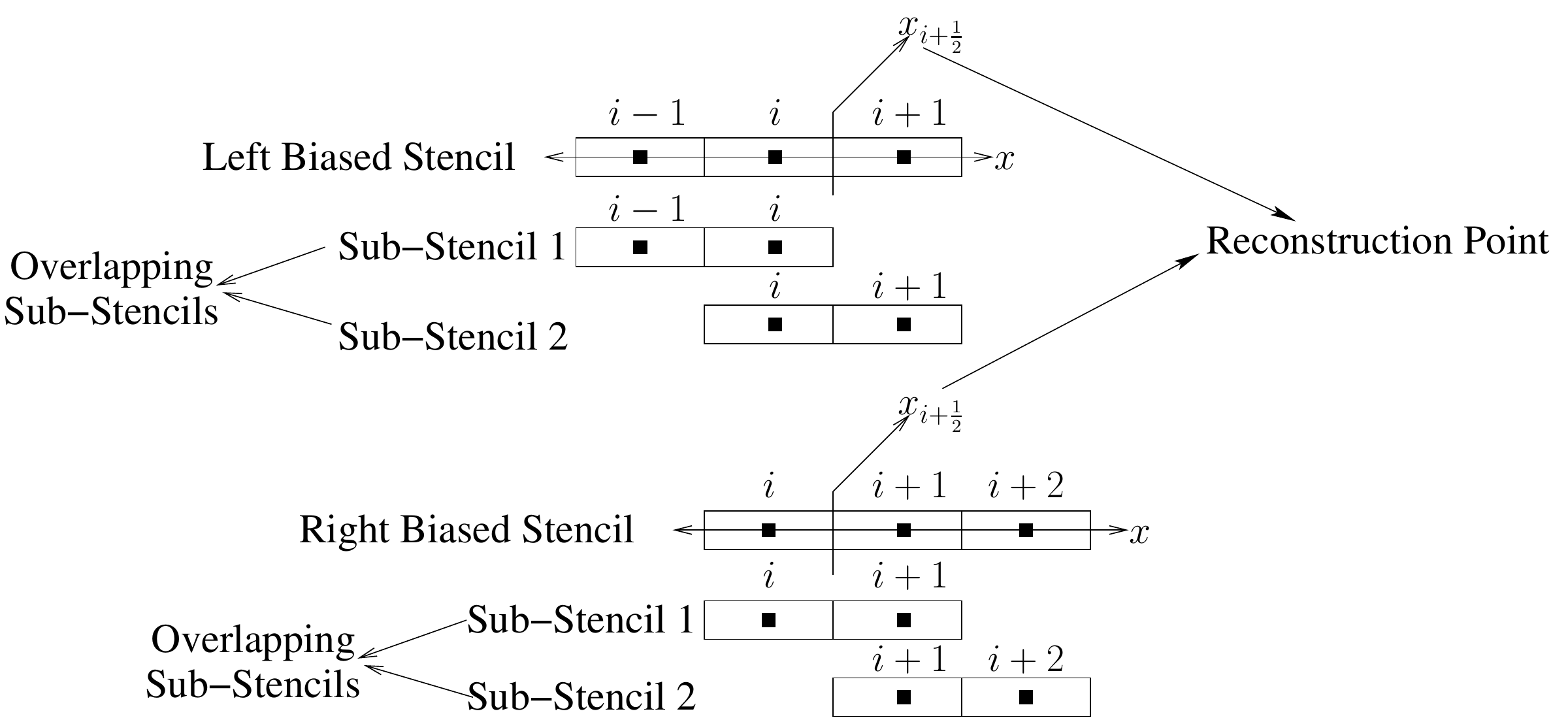}
  \caption{WENO-NP3, choice of Stencils and sub-stencils. Illustrates bias with respect to the point of reconstruction, which is $x_{i+\frac{1}{2}}$.}
  \label{fig:W3Stencils}
\end{center}
\end{figure}
For WENO-NP3 reconstruction, a stencil of 3 points is used (see figure~\ref{fig:W3Stencils} for stencils and sub stencils).
Equation~(\ref{eq:discreteHypConsLawFS}) is used to advance from time $t_n$ to $t_{n+1}$.  At grid point
$x_i$, approximations $\hat{h}^{\pm}_{i+\frac{1}{2}}$ and $\hat{h}^{\pm}_{i-\frac{1}{2}}$ (subscript $n$, indicating time level, is dropped for brevity) to
$h^{\pm}(x_{i+\frac{1}{2}}, t_n)$ and $h^{\pm}(x_{i-\frac{1}{2}}, t_n)$, respectively, are needed. These are given by the following equations:
\begin{equation}
  \hat{h}^{\pm}_{i+\frac{1}{2}} = \omega^{\pm}_{1}H^{\pm}_1 + \omega^{\pm}_{2}H^{\pm}_2
\end{equation}
Formulae for $H^{\pm}_1, H^{\pm}_2, \omega^{\pm}_1 \text{, and } \omega^{\pm}_2$ are given below,
\begin{align}
  H^{+}_1 = \frac{3E^{+}_i - E^{+}_{i-1}}{2},~ & H^{+}_2 = \frac{E^{+}_i + E^{+}_{i+1}}{2},\\
  H^{-}_1 = \frac{E^{-}_i + E^{-}_{i+1}}{2},~ & H^{-}_2 = \frac{3E^{-}_{i+1} - E^{-}_{i+2}}{2},\\
\omega^{\pm}_j = \frac{\tilde{\omega}^{\pm}_j}{\bar{\omega}^{\pm}},~& \bar{\omega}^{\pm} = \tilde{\omega}^{\pm}_1 + \tilde{\omega}^{\pm}_2, \\
  \tilde{\omega}^{\pm}_j = \gamma^{\pm}_j\Bigg(1 +  \frac{\tau^{\pm}_{NP}}{(\beta^{\pm}_j+\epsilon)}\Bigg),~& \epsilon = 10^{-14}, \text{ for }j=1, 2 ,\\
  \gamma^{+}_1 = \frac{1}{3},~ & \gamma^{+}_2 = \frac{2}{3},\\
  \gamma^{-}_1 = \frac{2}{3},~ & \gamma^{-}_2 = \frac{1}{3},\\
  \beta^{+}_1 = (E^+_{i-1} - E^+_i)^2,~ & \beta^{+}_2 = (E^+_{i+1} - E^+_i)^2,\\
  \beta^{-}_1 = (E^-_{i+1} - E^-_i)^2,~ & \beta^{-}_2 = (E^-_{i+1} - E^-_{i+2})^2,\\
  \tau^{\pm}_{NP} = \Bigg|\beta^{\pm}_3 - & ~\frac{\beta^{\pm}_1 + \beta^{\pm}_2}{2}\Bigg|^{1.5}\\
  \beta^{+}_3 = \frac{1}{4}(E^+_{i-1} - E^+_{i+1})^2 + &\frac{13}{12}(E^+_{i-1} - 2E^+_{i} + E^+_{i+1})^2,\\
  \beta^{-}_3 = \frac{1}{4}(E^-_{i} - E^-_{i+2})^2 + &\frac{13}{12}(E^-_{i} - 2E^-_{i+1} + E^-_{i+2})^2,\\
  E^{\pm}_k = E^{\pm}(x_k, t_n)&\text{, for }k=i-1, i, i+1, i+2.
\end{align}

$H^{\pm}_1$, $H^{\pm}_2$ are lower order approximations to $h^{\pm}(x_{i+\frac{1}{2}}, t_n)$ and are calculated using the relevant sub-stencils shown in figure~\ref{fig:W3Stencils}.
The linear weights $\gamma^{\pm}_1$, $\gamma^{\pm}_2$ and the smoothness indicators $\beta^{\pm}_1$, $\beta^{\pm}_2$, and $\beta^{\pm}_3$ are used to calculate the nonlinear weights $\omega^{\pm}_1$,
$\omega^{\pm}_2$. A convex combination of $H^{\pm}_1$, $H^{\pm}_2$, with the corresponding nonlinear weights is taken to obtain the final weighted essentially non-oscillatory
reconstruction.

This procedure is for spatial discretisation of hyperbolic conservation law in one space dimension. For equations in two space dimensions such as,
\begin{equation}
  \frac{\partial }{\partial t}Q(x, y, t) + \frac{\partial}{\partial x}E(Q(x, y, t)) + \frac{\partial}{\partial y}F(Q(x, y, t)) = 0,
\end{equation}
the same procedure can be used for discretising the $x$ and $y$ derivatives separately.
The resulting semi-discrete form is integrated in time using TVD-RK3 method, which is described below.

\subsection{TVD-RK3 time discretisation}
Consider the equation
\begin{equation}
  \label{eq:timInt}
  \frac{d}{dt} u = L(u).
\end{equation}
The simple forward Euler time discretisation between two time levels $t_n$ and $t_{n+1}$ separated by $\Delta t$ is given by
\begin{equation}
  \label{eq:fowEulDisc}
  u^{n+1} = u^{n} + \Delta t L(u^{n}).
\end{equation}
A three stage third order TVD (Total Variation Diminishing) or SSP (Strong Stability Preserving) \cite{gottlieb2001} Runge-Kutta discretisation is given by
\begin{align}
  u^{(1)} &= u^{n} + \Delta t L(u^{n}),\\
  u^{(2)} &= \frac{3}{4} u^n + \frac{1}{4}u^{(1)} + \frac{1}{4}\Delta t L(u^{(1)}),\\
  u^{n+1} &= \frac{1}{3} u^n + \frac{2}{3}u^{(1)} + \frac{2}{3}\Delta t L(u^{(2)}).
\end{align}

The TVD-RK3 discretisation is used to advance in time from $t_n$ to $t_{n+1}$.
Next, the standard symmetry technique used for applying wall boundary conditions is described.
\subsection{Governing equations and the standard symmetry technique}\label{Sec:SymmCondAndProbs}
We start with the two-dimensional Euler equations, which are
\begin{equation}
  \label{eq:Euler2dDiff}
  \frac{\partial Q}{\partial t} + \frac{\partial E}{\partial x} + \frac{\partial F}{\partial y} = 0
\end{equation}
where
\begin{equation}
  \label{eq:Euler2dDiffTerms}
~Q = \begin{bmatrix}\rho \\ \rho u \\ \rho v \\ \rho e_t\end{bmatrix}, ~E =  \begin{bmatrix}\rho u \\ \rho u^2 + p \\ \rho uv \\ (\rho e_t + p)u\end{bmatrix},
~F =  \begin{bmatrix}\rho v \\ \rho vu \\ \rho v^2 +p \\ (\rho e_t + p)v\end{bmatrix},
 ~ e_t = \frac{p}{\rho(\gamma -1)} + \frac{1}{2}\left(u^2 + v^2 \right)
\end{equation}

\begin{figure}[!htbp]
  \centering
  \begin{subfigure}{.5\textwidth}
    \centering
  \includegraphics[height=.23\textheight]{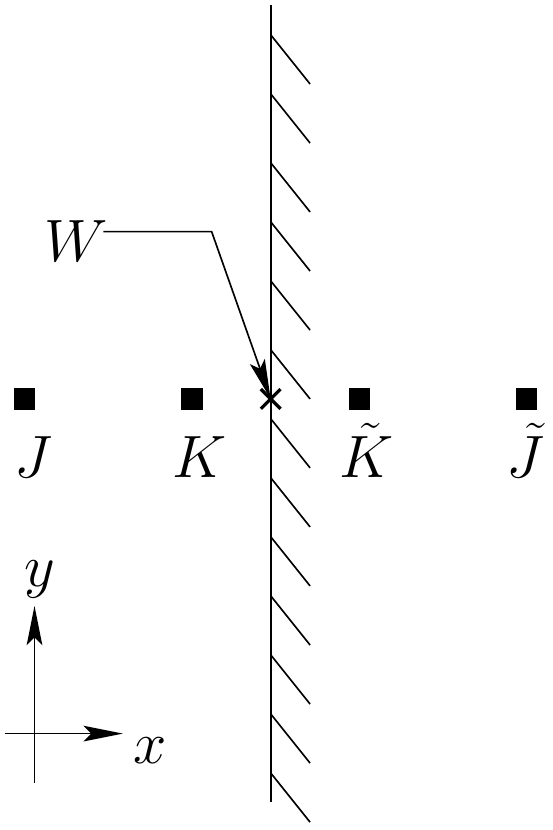}
  \caption{MNGW: No Grid point at W.}
  \label{fig:sSTMNGW}
\end{subfigure}%
\begin{subfigure}{.5\textwidth}
  \centering
  \includegraphics[height=.23\textheight]{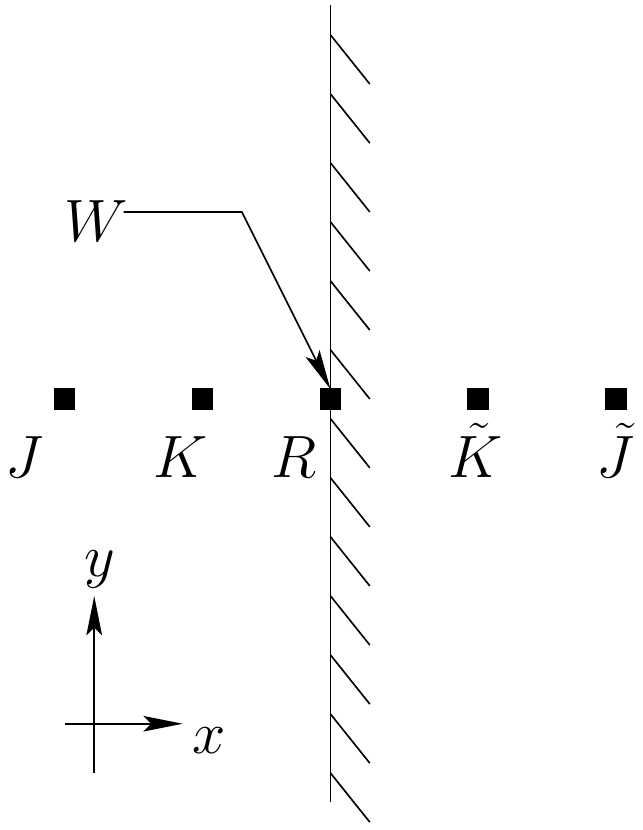}
  \caption{MGW (Has grid points on wall - R)}
  \label{fig:sSTMGW}
\end{subfigure}
\caption{Free slip wall: $\tilde{J}$, $\tilde{K}$ are ghost points placed so as to be located symmetrically with respect to the wall corresponding to physical grid points $J$ and $K$, respectively. State at $\tilde{K}$ and $\tilde{J}$ are obtained using equations in table~\ref{tab:SSTEquations}.}
\label{fig:wallReflectionBC}
\end{figure}

In finite difference schemes, wall boundary conditions can be applied
by using ghost points (see Figure
\ref{fig:wallReflectionBC}). 
We classify meshes into two types, mesh with no grid points on the wall (MNGW - see figure~\ref{fig:sSTMNGW})  and mesh with grid points on the wall (MGW - see figure
\ref{fig:sSTMGW}). 
For meshes with grid point on the wall, the state on the wall boundary is available either to directly apply the boundary condition, or, to verify if
the applied boundary condition is producing the appropriate state on the wall. For a mesh without grid point on the wall, the boundary condition can be applied using the ghost
points but the state at the boundary is not directly available and must be inferred.
%1. Therefore state on the wall boundary is not available directly and checking if boundary condition is satisfied there is problematic.\\
%2. Therefore it is not possible to prescribe the state on the boundary accurately.\\
%3. Consequently modelling or capturing the wall boundary as accurately as when there are grid points on the wall is not possible. Also, the state on the wall boundary is
%not available directly. Therefore checking if the boundary condition is satisfied there is problematic.\\
%4. Consequently asserting that the wall boundary is captured as accurately as when there are grid points on the wall is not possible. This is because the state
%on the wall boundary is not available and therefore checking if the boundary condition is satisfied there is problematic.

The application of the standard symmetry technique at point W (see figure~\ref{fig:wallReflectionBC}) involves setting the states at ghost grid points $\tilde{J}$ and $\tilde{K}$, which are located symmetrically
with respect to the wall corresponding to grid points $J$ and $K$, respectively. The previous two sections have shown how the states at J, K and R are determined. The state
at $\tilde{K}$
is found using the state at grid point $K$ and the equations in table~\ref{tab:SSTEquations} (subscripts of flow properties are used to indicate grid points).
%>{\centering\arraybackslash}m{0.2\textwidth}
\begin{table}[!htbp]
  \begin{center}
  \caption{Standard symmetry technique equations and corresponding boundary condition approximated at the wall boundary
    (see figure~\ref{fig:wallReflectionBC} for grid point labels). Subscripts are used to indicate grid point. }
\label{tab:SSTEquations}
%      \renewcommand{\arraystretch}{1.25}
%    \begin{tabular}{|p{0.05\textwidth}|p{0.28\textwidth}|p{0.38\textwidth}|}
    \begin{tabular}{|>{\centering\arraybackslash}m{0.05\textwidth}|>{\centering\arraybackslash}m{0.2\textwidth}|>{\centering\arraybackslash}m{0.5\textwidth}|}
      \hline
      S.No. & SST Equations & Boundary condition that is approximated at point W \\ \hline
      1 & $p_{\tilde{K}} = p_{K}$ & $\partial p/\partial x = 0$ \\ \hline
      2 & $\rho_{\tilde{K}} = \rho_K$ & $\partial \rho/\partial x = 0$ \\ \hline
      3 & $(\rho u)_{\tilde{K}} = -(\rho u)_K$ & $\rho u = 0$ \\ \hline
      4 & $(\rho v)_{\tilde{K}} = (\rho v)_K$ & $\partial (\rho v)/\partial x = 0$ \\ \hline
    \end{tabular}
  \end{center}
\end{table}
%In the table (\ref{tab:SSTEquations}) $p_{\tilde{K}}$ is the pressure at the point $\tilde{K}$.
%The subscripts in the table are used to indicate grid point.
The negation of normal component of momentum density in the ghost points ensures a zero normal component of velocity on the wall. The symmetry of tangential momentum will allow slip.
The same process is used for other ghost points like $\tilde{J}$.

For code verification and validation, the test problems of isentropic vortex moving in uniform flow \cite{shu1998} and the problem of a shock reflection off a flat
plate \cite{yee1985} are used.

\subsection{Isentropic Vortex moving in uniform flow}
The initial condition for this problem is an isentropic vortex perturbation added to a uniform flow in the positive $x$ direction. The solution at any time $t$ is given by:
\begin{align}
  u(x, y, t) &= u_0 - \beta e^{(1-r^{2})} \frac{y-y_{0}}{2\pi},\\
  v(x, y, t) &= \beta e^{(1-r^{2})} \frac{x-x_{0}-u_0 t}{2\pi},\\
  \rho(x, y, t) &= \left(1 - \left(\frac{\gamma - 1}{16\gamma \pi^{2}}\right)\beta^{2} e^{2(1-r^{2})}\right)^{\frac{1}{\gamma-1}},
\end{align}
where $p(x, y, t) = (\rho(x, y, t))^{\gamma}$ and $r=\sqrt{(x-x_{0}-u_0 t)^{2}+(y-y_{0})^{2}}$. The parameter values chosen are $x_{0}=8$, $y_{0}=0$, $\beta=2.0$, $u_0 = 1.0$
and $\gamma = 1.4$. The computational domain is a square of dimensions $16 \text{units} \times 16 \text{units}$ with $0 \leq x \leq 16$ and $-8 \leq y \leq 8$. Periodic
boundary conditions are applied along the $x$ and $y$ directions. Using the properties at $t=0$ as initial conditions, the numerical method
described before is used to obtain a solution at $t=2.34375$ units. The time step ${\Delta t=0.09765625 \Delta x}$. This problem was run for meshes with grid point
spacings ($GPS=\Delta x = \Delta y$) of $1/25, 1/50, 1/75, 1/100, 1/150, 1/175, 1/200, \text{ and } 1/225$. The $L_1$ and
$L_{\infty}$ errors for meshes with different GPS and the observed order of accuracy are given in table~\ref{tab:L1LInfErrors}.

\begin{table}[!htbp]
  \begin{center}
  \caption{$L_1$ and $L_{\infty}$ errors of total energy density ($\rho e_t$) for different GPS for the problem of isentropic vortex moving in a uniform flow and observed order of accuracy.}
\label{tab:L1LInfErrors}
  \begin{tabular}{|>{\centering\arraybackslash}m{0.05\textwidth}|>{\centering\arraybackslash}m{0.14\textwidth}|>{\centering\arraybackslash}m{0.08\textwidth}|>{\centering\arraybackslash}m{0.14\textwidth}|>{\centering\arraybackslash}m{0.08\textwidth}|}
      \hline
      GPS  &$L_1\text{ error}$ $\times 10^8$ & $L_1\text{ order}$ & $L_{\infty}\text{ error}$ $\times 10^6$ & $L_{\infty}\text{ order}$ \\ \hline 
      1/25 & 862.397 & - & 1117.663 & - \\ \hline
      1/50 & 105.379 & 3.03 & 156.781 & 2.83 \\ \hline
      1/75 & 31.007 & 3.02 & 78.321 & 1.71 \\ \hline
      1/100 & 13.041 & 3.01 & 24.300 & 4.07 \\ \hline
      1/150 & 3.864 & 3.00 & 9.611 & 2.29 \\ \hline
      1/200 & 1.630 & 3.00 & 4.601 & 2.56 \\ \hline
      1/225 & 1.144 & 3.00 & 2.965 & 3.73 \\ \hline
    \end{tabular}
  \end{center}
\end{table}

\subsection{Shock reflection off a flat plate}
\begin{figure}[!htbp]
\begin{center}
  \includegraphics[width=0.8\textwidth]{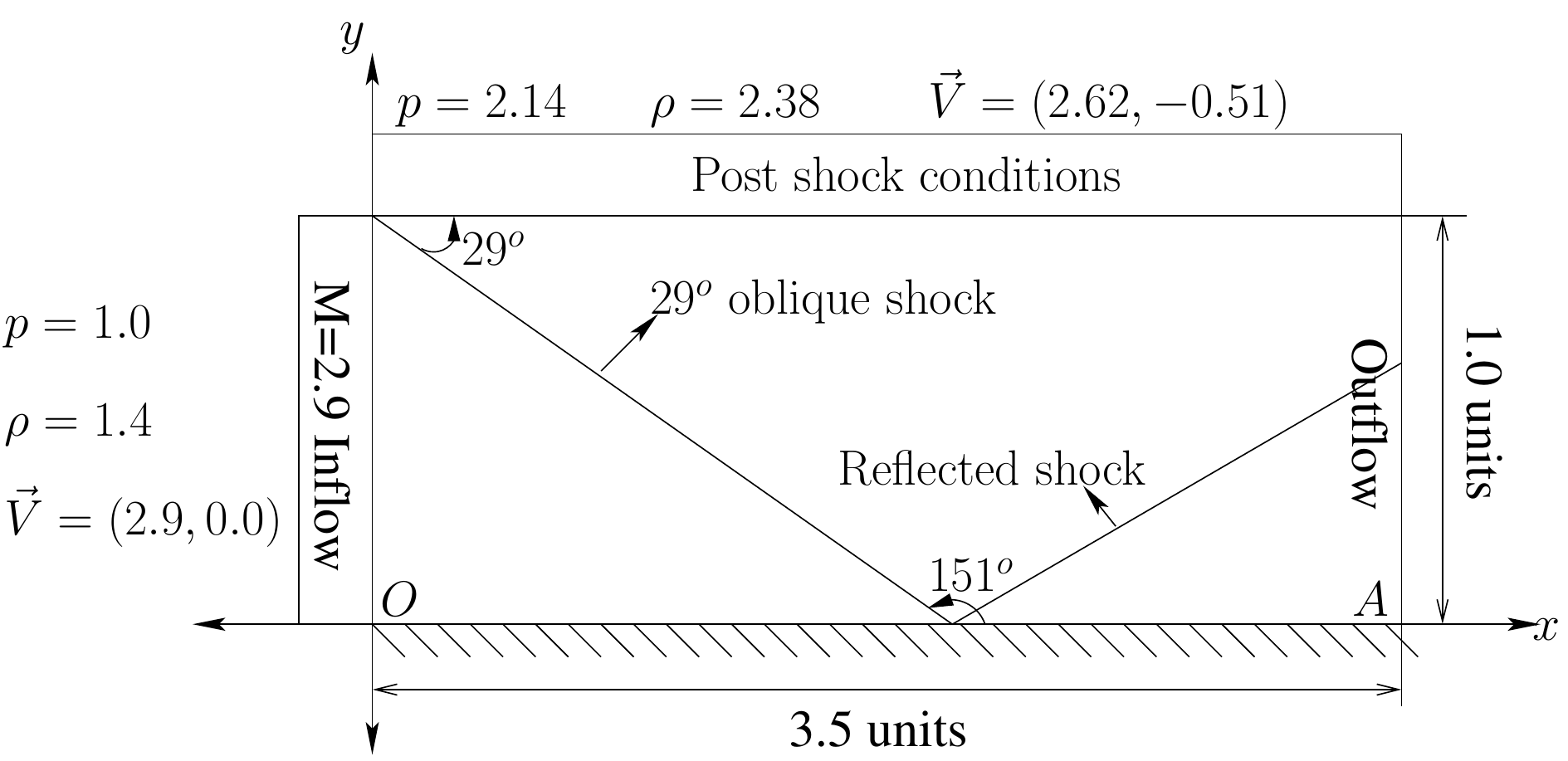}
  \caption{Shock reflection off a flat plate (not drawn to scale): Sketch of problem domain and flow conditions. OA is a solid wall and SST is used to apply wall boundary conditions.}
  \label{fig:shockRefGeo}
\end{center}
\end{figure}

\begin{figure}[!htbp]
\begin{center}
  \includegraphics[width=0.9\textwidth]{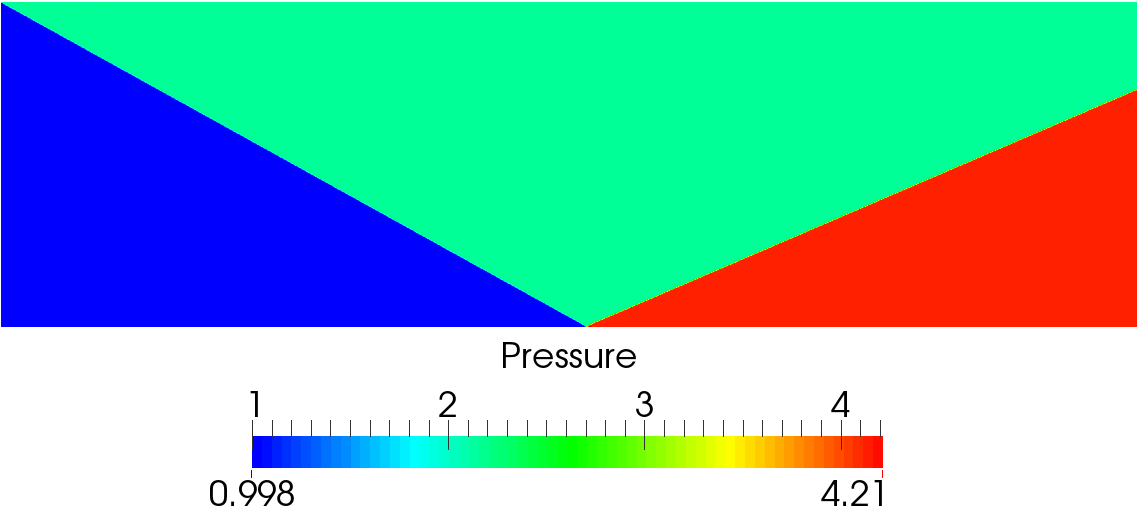}
  \caption{Color plot of pressure for the problem of 29\textdegree{} Mach 2.9 shock reflection off a flat plate obtained using MGW with GPS of $1/800$}
  \label{fig:shockRefPres}
\end{center}
\end{figure}

The problem of shock reflection off a flat plate \cite{yee1985} is used for validating SST and the numerical scheme.
The governing differential equations are given by  equation~(\ref{eq:Euler2dDiff}). The computational domain along with the boundary conditions are shown in figure
\ref{fig:shockRefGeo}. A shock, with shock wave angle of 29\textdegree{}, pre-shock Mach
number of 2.9 reflects off a flat plate. The computational domain is a rectangle of size 3.5 units by 1.0 units in the $x$ and $y$ directions, respectively. The bottom boundary,
`$y=0$', is a free slip wall. The left boundary, `$x=0$' is an inflow with $M=2.9$ flow in the positive $x$-direction, $p = 1.0$, $\rho = 1.4$. On and above the top boundary
`$y=1.0$', post oblique shock state is prescribed. The shock makes an angle of 151\textdegree{} with the $x$ axis, as shown in figure~\ref{fig:shockRefGeo}. The flow field is
initialized with a Mach 2.9 flow with velocity vector pointed in the positive $x$-direction. 
A CFL number of $0.125$ was chosen to calculate the global time step value. The `$\alpha$' in the Lax-Freidrichs flux splitting is given by
$\alpha = (2.9+1.0) = 3.9$. Computations were done using MGW with grid point spacings or GPS ($= \Delta x = \Delta y$) of $1/50, 1/100$, $1/200$, $1/400$, and $1/800$.
Color plot of pressure obtained using MGW with GPS of $1/800$ is shown in figure~\ref{fig:shockRefPres}.

The values of pressure, post the incident and reflected shock, of 2.14 and 4.10, obtained from the numerical solution are in good agreement with those of the analytic
solution. The reflected shock in
the numerical solution makes an angle of approximately 23\textdegree{} with the $x$-axis, which is also in good agreement with the analytic solution. Having tested the solver
along with the standard symmetry technique, we turn to the problem of walls with corners.

%As mentioned earlier, walls with expansion corners have two problems. One, determining the state at the corner is not possible. Two, it will be shown that using the SST will lead
%to mass leak through the wall near the corner. This will be demonstrated using the problem of Mach 4.0 flow over a forward facing step in the next section.

\section{Demonstration of flow leak near a corner}\label{Sec:DemonstrateLeak}
In this section, we demonstrate flow leak near the corner using the problem of Mach 4.0 flow over a forward facing step and we begin with the problem definition.

\subsection{Mach 4.0 flow over a forward facing step: Problem domain and boundary conditions}\label{Sec:FowFacStepGeo}
\begin{figure}[!htbp]
\begin{center}
  \includegraphics[width=.65\textwidth]{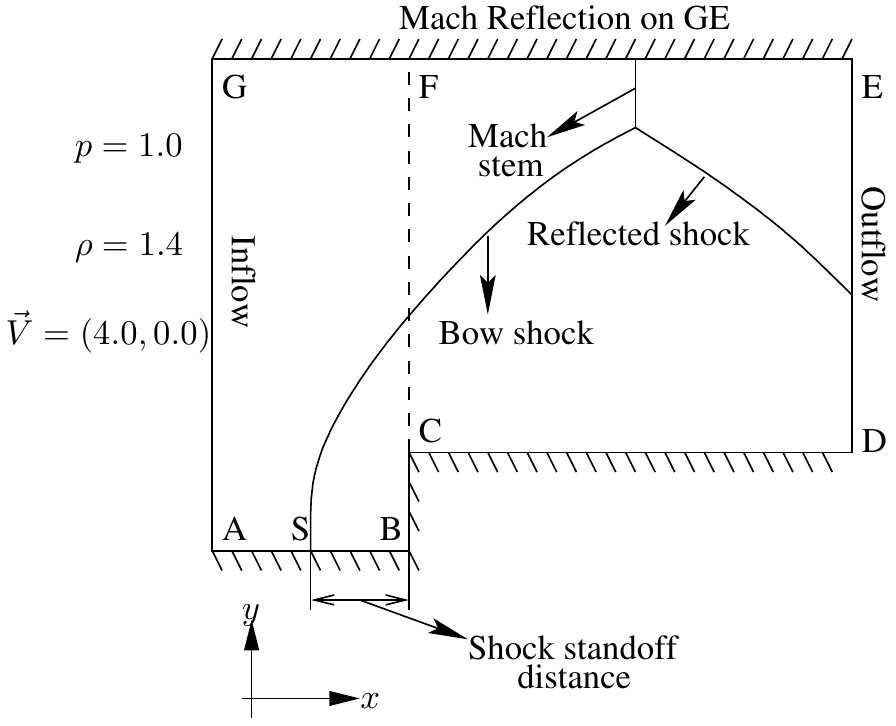}
  \caption{Forward facing step: Sketch of problem domain, boundary conditions and the expected shock structure (not drawn to scale).}
  \label{fig:fowFacStepGeo}
\end{center}
\end{figure}
Figure~\ref{fig:fowFacStepGeo} shows the geometry of the flow field. Except for the inflow and outflow, all of the boundaries are free slip walls. In non-dimensional units
$AG = 1.0$, BC, the step height which should be $20\%$ of $AG$, is equal to $0.2$ and AB $= 0.6$. The inflow is supersonic and the state of Mach 4.0 flow in the positive
$x$ direction with
$\rho = 1.4$, $p = 1.0$, is prescribed there. For ease of applying boundary conditions, CD $= 0.6$ is chosen so that the outflow will be
supersonic. One sided differences, biased in the negative $x$ direction are used to calculate $x$ derivatives near the outflow. The inflow conditions are prescribed at
all grid points as the initial conditions. For those conditions, speed of sound $a=1$ and $\alpha\text{ (for Lax-Freidrichs flux splitting)} = 5.0~(=a+M)$.
%which is inflow Mach number of 4.0, with the velocity vector in the positive x direction, $\rho = 1.4$, $p = 1.0$, which makes
%speed of sound $a=1$ and $\alpha =5.0$($=1.0+M$).
As mentioned earlier, state at the expansion corner point `C' (see figure~\ref{fig:fowFacStepGeo}) can not be determined and ways of tackling that problem are given next.

\subsection{Tackling the `corner point state' problem}\label{Sec:tacklingCornerPt}

As mentioned earlier determining the state at the corner is not possible.
%One way of avoiding this problem
%is by approximating the corner with an arc of a circle with a small radius of curvature. As this study is restricted to Cartesian meshes, this strategy is not employed.
\begin{figure}[!htbp]
  \centering
  \begin{subfigure}{.4\textwidth}
    \centering
  \includegraphics[height=0.2\textheight]{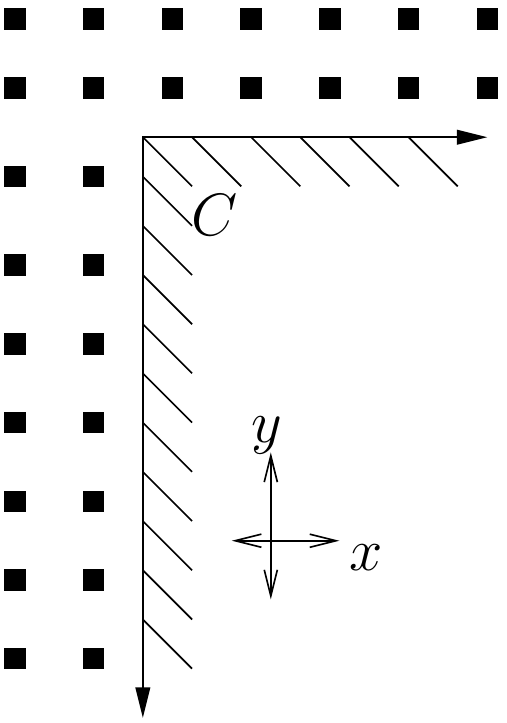}
  \caption{MNGW: No grid point at corner}
  \label{fig:noGrPtOnCorner}
\end{subfigure}%
\begin{subfigure}{.4\textwidth}
  \centering
  \includegraphics[height=0.2\textheight]{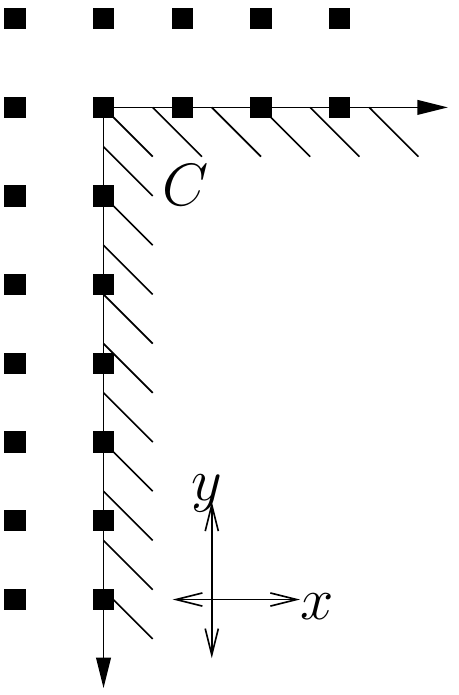}
  \caption{MGW: Has grid point at corner}
  \label{fig:grPtOnCorner}
\end{subfigure}
  \caption{Two Cartesian mesh configurations near corner}
  \label{fig:withWithoutCornGrPt}
\end{figure}
One way to avoid this problem is to choose the mesh such that there is no grid point at the corner, but this results in the MNGW grid as shown in figure 
\ref{fig:noGrPtOnCorner} and a corresponding approximate application of boundary condition.

\begin{figure}[!htbp]
\begin{center}
  \includegraphics[width=0.45\textwidth]{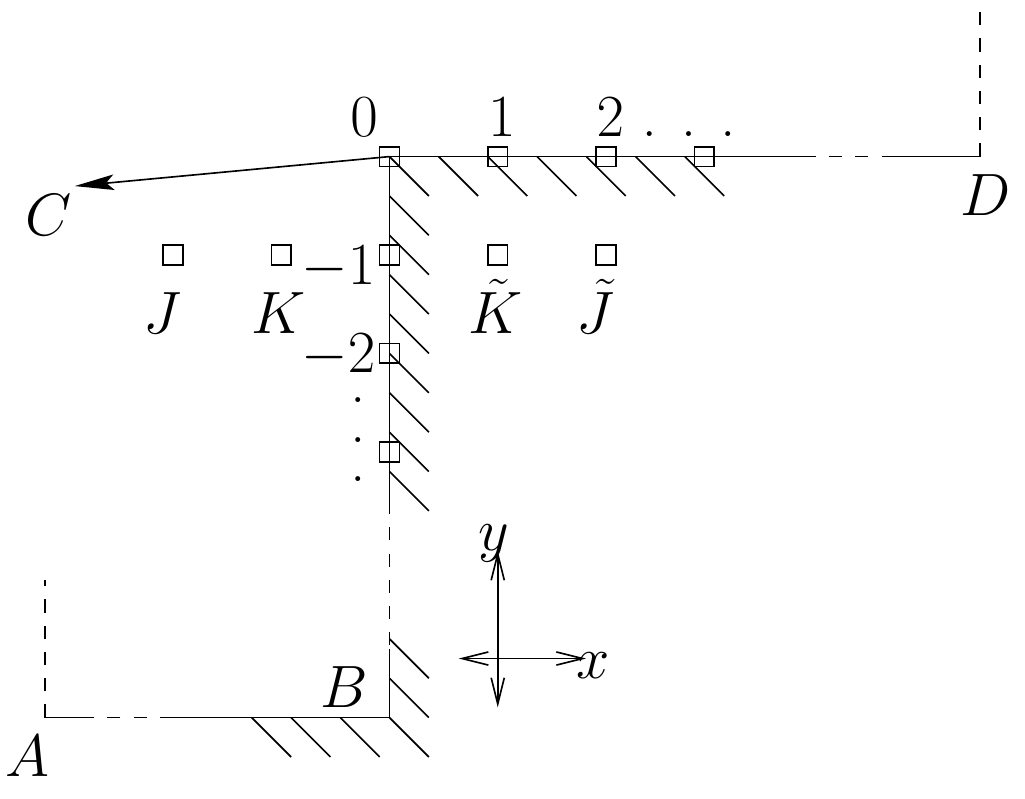}
  \caption{Mesh with grid point at corner (MGW). Sample interior grid points $J$ and $K$ are shown along with corresponding ghost points. On the wall, grid points are 
    labelled ... -2, -1, 0, 1, 2 ..., which are diagnostic points and are used later for plotting.}
  \label{fig:gridPointAtCornerLab}
\end{center}
\end{figure}
To avoid this, we can have a mesh with grid points on the wall (MGW) as shown in figure~\ref{fig:grPtOnCorner} or \ref{fig:gridPointAtCornerLab}, which will also have a
grid point at the corner. Now, the problem of
determining state at grid point `$0$' needs to be addressed. Problems with corners whilst solving other partial differential equations
were addressed in \cite{holsteinJr1981, acheUnivWisconson1987, verhoff2004}. 
%As mentioned earlier approximate
%analytic solutions for the compressible, inviscid, subsonic flow over a ramp were published by Verhoff \cite{verhoff2004}, the first of which predicted infinite momentum 
%density at the corner. A fix was proposed to tackle that singularity and assign a state at the corner at steady state.
For the Euler equations on Cartesian mesh, we propose the following fix for the `corner point state problem'.
In a Cartesian mesh, unlike compression corner points (like `$B$' in figure~\ref{fig:gridPointAtCornerLab}), expansion corner points (like `$0$' in
figure~\ref{fig:gridPointAtCornerLab}) have all the necessary grid points required to 
discretise the Euler equations with an appropriate upwind biasing to the full order of the scheme. Therefore, we propose solving the governing differential equation at the
corner, instead of applying boundary conditions.
To repeat, no boundary condition is applied at `$0$'. This fix will be referred to as `corner fix'. The point `0'
becomes an interior point and the corner is, in a sense, `rounded'.

Using the corner fix implies that the no-penetration condition is violated in the portion of the boundary between grid points $-1$, $0$ and also between $0$, $1$
(see figure~\ref{fig:gridPointAtCornerLab}). This is because using the corner fix will lead to a finite non-zero velocity at the corner grid point $0$.
Assigning any state with non-zero velocity at grid point $0$ will lead to this violation.
In addition to this, there will be flow leak in the region between grid points `$1$' - $D$ and also between grid points `$-1$' - $B$.
This leak will be demonstrated using the solution for the Mach 4.0 flow over a forward facing step obtained using MGW and the `corner fix', employing SST for applying wall
boundary conditions.
%This numerical solution has a flow leak through the wall boundary near the expansion corner point C ( see figure \ref{fig:fowFacStepGeo}). Next, this flow leak will be analysed and its 
%causes will be explained subsequently.
%%It will be shown that using this modified symmetry technique (MST)
%%leads to prediction of grid independent shock structure even for coarse meshes with GPS = $1/50$.

\subsection{Flow leak near corner}\label{Sec:FowFacStepSymmCondsNoPen}
%\begin{figure}[!htbp]
%\begin{center}
%  \begin{subfigure}{0.48\textwidth}
%    \centering
%    \includegraphics[width=0.9\textwidth]{images/M40DX1By100/CC1OfCC1CC2Comp.png}
%    \caption{Solution obtained using MNGW, shock standoff distance = $0.27$ units}
%  \end{subfigure}%
%  \begin{subfigure}{0.48\textwidth}
%    \centering
%    \includegraphics[width=0.9\textwidth]{images/M40DX1By100/CC2OfCC1CC2Comp.png}
%    \caption{Solution obtained unsing MGW, shock standoff distance = $0.29$ units. }
%  \end{subfigure}
%  \begin{subfigure}{0.9\textwidth}
%    \centering
%    \includegraphics[width=0.9\textwidth]{images/M40DX1By100/landscapeLegendOfCC1CC2Comp.png}
%    \caption{Color map}
%  \end{subfigure}
%  \caption{Color plot of density, with shock standoff distance marked, for $M=4.0$ flow over a step(ABCFG portion), using SST for MNGW and MGW, with GPS = $\frac{1}{100}$.}
%  \label{fig:comparWithAndWithoutCornGrPt}
%\end{center}
%\end{figure}

%\begin{figure}[!htbp]
%  \begin{center}
%    \def\svgwidth{0.8\columnwidth}
%    \input{images/leakM30D10DX1By100.pdf_tex}
%    \caption{$M=4.0$ flow over forward facing step, plot of mass and energy flux normal to the wall boundary, near corner, due to using Standard Symmetry Technique, grid point spacing = $1/100$, vs grid point index(indexing as shown in figure \ref{fig:gridPointAtCornerLab}).}
%\label{fig:flowLeakPlot}
%\end{center}
%\end{figure}
Numerical solution to the problem of Mach 4.0 flow over a step, described in section~\ref{Sec:FowFacStepGeo}, is obtained using the numerical method described in
sections~\ref{Sec:shuOsherConsFinDiffSch} - \ref{Sec:SymmCondAndProbs}.
\begin{figure}[!htbp]
  \begin{center}
    \includegraphics[width=0.8\textwidth]{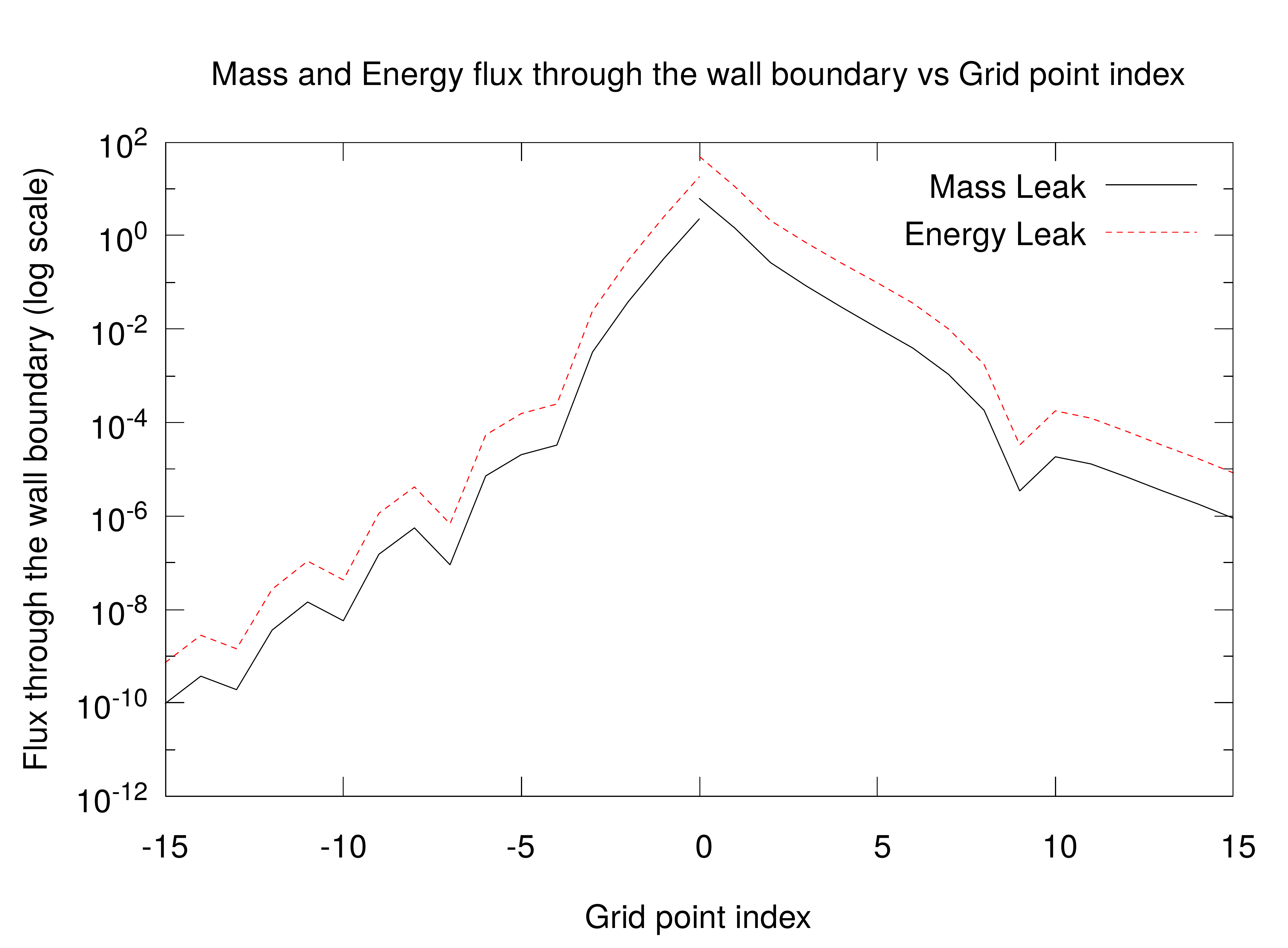}
    \caption{Mach 4.0 flow over forward facing step, plots of mass and energy flux through the wall boundary, near corner, due to using Standard Symmetry Technique vs grid point index (indexing as shown in figure~\ref{fig:gridPointAtCornerLab}), for grid point spacing of $1/100$.}
\label{fig:flowLeakPlot}
\end{center}
\end{figure}
%Figure \ref{fig:comparWithAndWithoutCornGrPt}, shows a comparison of color plots of density obtained using mesh MNGW and MGW
%for a grid point spacing
%of $\frac{1}{100}$. These plots look essentially same, however, the shock standoff distance measured along the bottom wall obtained using MNGW and MGW is $0.27$, $0.29$
%units, respectively.
%%try to plot streamlines or velocity vectors to show the leak
The problem with the numerical solution obtained is that there is flow leak through the wall near the corner.
Figure~\ref{fig:flowLeakPlot} has plot of magnitude of mass flux and energy flux through the wall boundary, below and downstream of the corner grid point. The grid points are
labelled as shown in figure
\ref{fig:gridPointAtCornerLab}. As can be seen, the no-penetration condition is violated for several grid points below and downstream of the corner point and hence there is non-zero mass flux and
energy flux through the wall boundary. This mass flux is a leak. It is integrated to find the total mass flow rate leaking, $\dot{m}_l$, near the corner. The total mass flow
rate into the domain through the inlet boundary AG (figure~\ref{fig:fowFacStepGeo}), is denoted by $\dot{m}_i$. Similarly, $\dot{e}_l$ and $\dot{e}_i$ can be defined for
the energy flow rate. The mass and energy
leak data in table~\ref{tab:M40massEnerLeakPercentage} is obtained by numerical integration (trapezoidal rule). The integration for leak below the corner is done from 
point 'B'  to grid point '-1' 
and that for downstream of corner is done from grid point `1'  to point 'D' (see figure~\ref{fig:gridPointAtCornerLab}). This total
mass (energy)
leak per unit time is expressed as a percentage of inflow rate of mass (energy) through face AG (see figure~\ref{fig:fowFacStepGeo}), in table
\ref{tab:M40massEnerLeakPercentage}. 
%This flow leak is not apparent in the solution obtained using mesh without grid point at the corner as there are no grid points on the wall.(want to say this?)

%    \begin{tabular}{|p{0.07\textwidth}|p{0.14\textwidth}|p{0.145\textwidth}|p{0.075\textwidth}|p{0.14\textwidth}|p{0.145\textwidth}|p{0.075\textwidth}|}
\begin{table}[!htbp]
  \begin{center}
  \caption{Leakage in a Mach 4.0 flow over forward facing step: $\dot{m}_l$ and $\dot{e}_l$ as a percentage of $\dot{m}_i$ and $\dot{e}_i$ (Positive value indicates that mass or energy is flowing in). }
\label{tab:M40massEnerLeakPercentage}
  \begin{tabular}{|>{\centering\arraybackslash}m{0.05\textwidth}|>{\centering\arraybackslash}m{0.13\textwidth}|>{\centering\arraybackslash}m{0.14\textwidth}|>{\centering\arraybackslash}m{0.09\textwidth}|>{\centering\arraybackslash}m{0.13\textwidth}|>{\centering\arraybackslash}m{0.14\textwidth}|>{\centering\arraybackslash}m{0.09\textwidth}|}
%    \begin{tabular}{|p{0.05\textwidth}|p{0.14\textwidth}|p{0.145\textwidth}|p{0.05\textwidth}|p{0.14\textwidth}|p{0.145\textwidth}|p{0.05\textwidth}|}
      \hline
      \multirow{2}{5em}{GPS}&\multicolumn{3}{c|}{Mass leak rate percentage ($\dot{m}_l/\dot{m}_i\times 100$)}&\multicolumn{3}{c|}{Energy leak rate percentage$(\dot{e}_l/\dot{e}_i\times 100$)}\\\cline{2-7}
                            &Below Corner ($a$)&Downstream of corner ($b$)&Total $(|a|+|b|)$&Below Corner ($c$)&Downstream of corner ($d$)&Total $(|c|+|d|)$\\ 
      \hline
      $1/50$ & -0.09 & 0.36 & 0.45 & -0.07 & 0.29 & 0.36 \\ \hline
      $1/100$ & -0.04 & 0.20 & 0.24 & -0.03 & 0.16 & 0.19 \\ \hline
      $1/200$ & -0.01 & 0.09 & 0.10 & -0.01 & 0.08 & 0.09 \\ \hline
    \end{tabular}
  \end{center}
\end{table}
The flow leak reduces with reduction in grid point spacing which can be asserted using data in table~\ref{tab:M40massEnerLeakPercentage}. Positive values indicate that mass or
energy
is flowing into the domain. The flow leaks out of the domain below the corner and leaks into the domain downstream of the corner as is evident from the percentages in table
\ref{tab:M40massEnerLeakPercentage}.

\begin{figure}[!htbp]
  \begin{center}
    \includegraphics[width=0.8\textwidth]{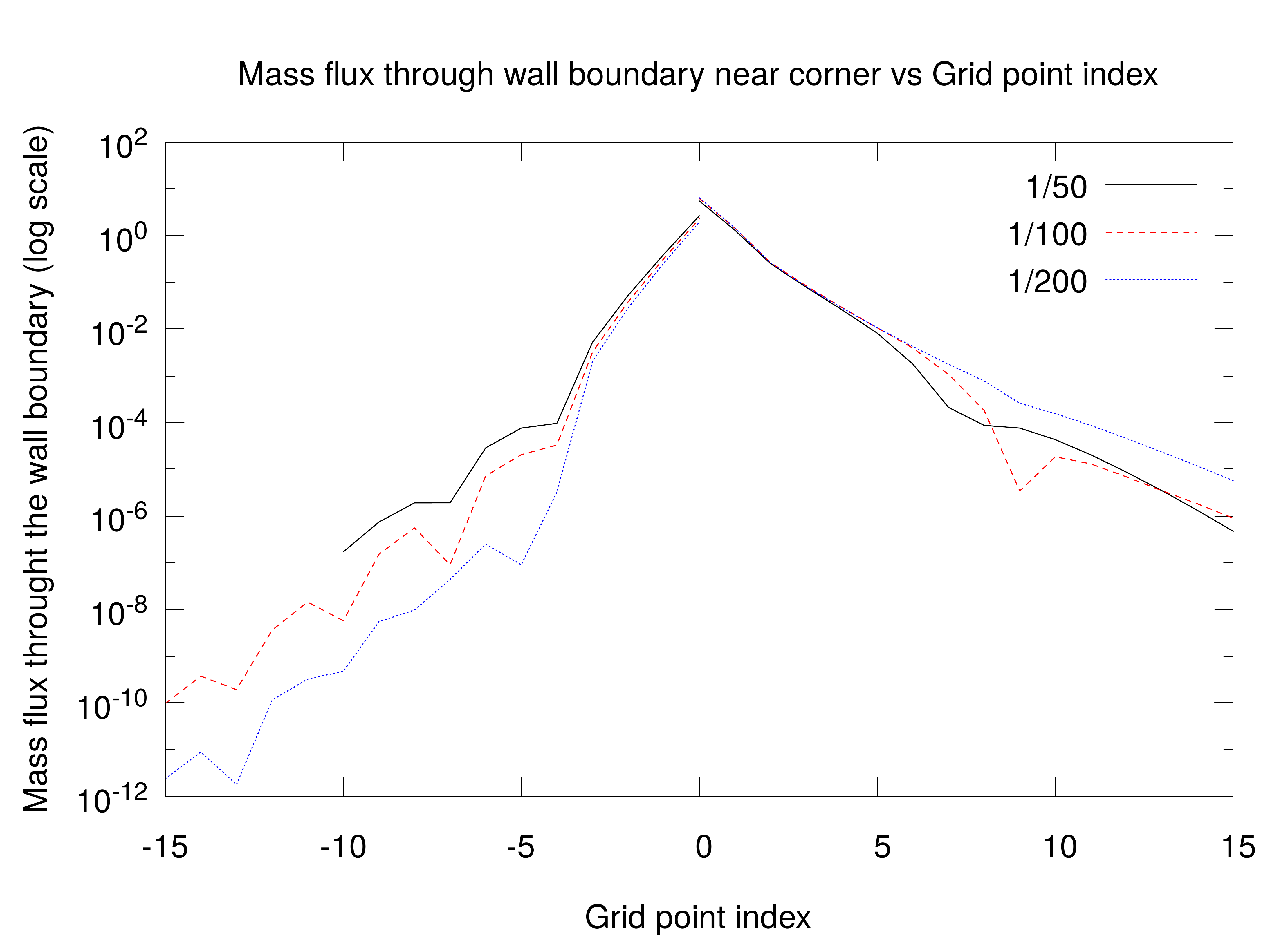}
    \caption{Mach 4.0 flow over forward facing step, plots of non-zero mass flux normal to the wall boundary, near corner, due to using SST vs grid point index (indexing as shown in figure~\ref{fig:gridPointAtCornerLab}), for GPS of $1/50$, $1/100$, $1/200$.}
\label{fig:massLeakComparM30}
\end{center}
\end{figure}
\begin{figure}[!htbp]
  \begin{center}
    \includegraphics[width=0.8\textwidth]{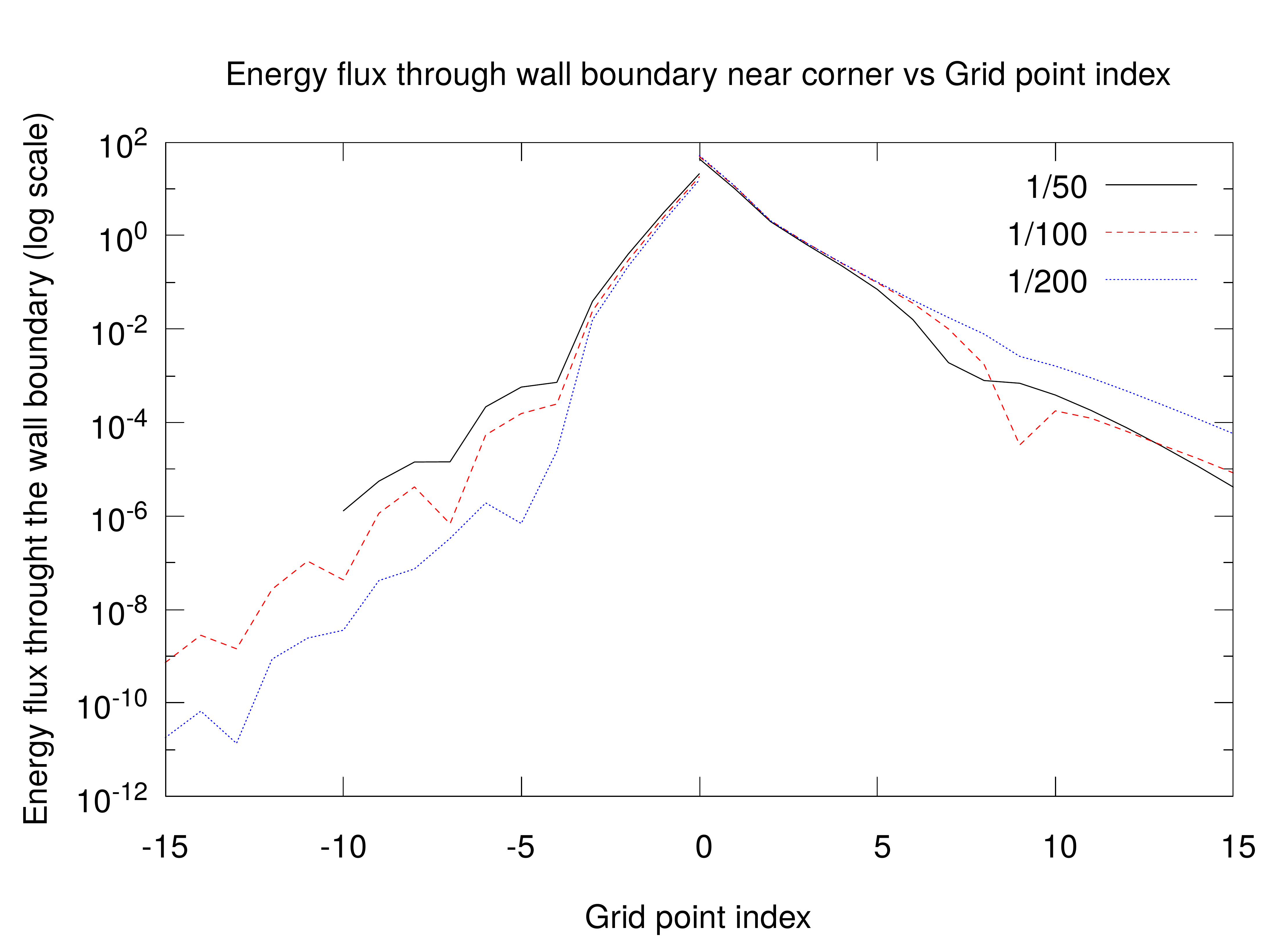}
    \caption{Mach 4.0 flow over forward facing step, plots of non-zero energy flux normal to the wall boundary, near corner, due to using SST vs grid point index (indexing as shown in figure~\ref{fig:gridPointAtCornerLab}), for GPS of $1/50$, $1/100$, $1/200$.}
\label{fig:enerLeakCompar}
\end{center}
\end{figure}
Figure~\ref{fig:massLeakComparM30} shows  plots of mass flux normal to the wall boundary near the corner for
GPS of $1/50$, $1/100$ and $1/200$. Figure~\ref{fig:enerLeakCompar} has the corresponding plot of energy flux. From the plots, it is evident that 
%the magnitude
%of the leak at a point is independent on the physical distance of the point from the corner but is essentially dependent on the number of grid points between a given point
%and the corner. Therefore 
the physical region over which flow leak happens decreases with decreasing grid point spacing.

The flow leak near the corner is due to the use of SST near the corner and this will be elaborated next.

%\begin{figure}[!htbp]
%\begin{center}
%  \begin{subfigure}{0.45\textwidth}
%    \centering
%    \includegraphics[width=0.9\textwidth]{images/M40DX1By100/CC2OfCC2CC3Comp.png}
%    \centering
%    \caption{MGW, SST, $0.29$, units is the shock\\ standoff distance.}
%  \end{subfigure}%
%  \begin{subfigure}{0.45\textwidth}
%    \centering
%    \includegraphics[width=0.9\textwidth]{images/M40DX1By100/CC3OfCC2CC3Comp.png}
%    \caption{MGW, SST with no-penetration enforced,\\ $0.27$ units is the shock standoff distance.}
%  \end{subfigure}
%  \begin{subfigure}{0.9\textwidth}
%    \centering
%    \includegraphics[width=0.9\textwidth]{images/M40DX1By100/landscapeLegendOfCC2CC3Comp.png}
%    \caption{Color map}
%  \end{subfigure}
%  \caption{Color plot of density and shock standoff distance for $M=4.0$ flow over a step(ABCFG portion), GPS = $\frac{1}{100}$.}
%  \label{fig:WithAndWithoutNoPen}
%\end{center}
%\end{figure}

\section{Analysis of standard symmetry technique near corners}\label{Sec:analOfSymmConds}

\begin{figure}[!htbp]
\begin{center}
  \includegraphics[width=0.45\textwidth]{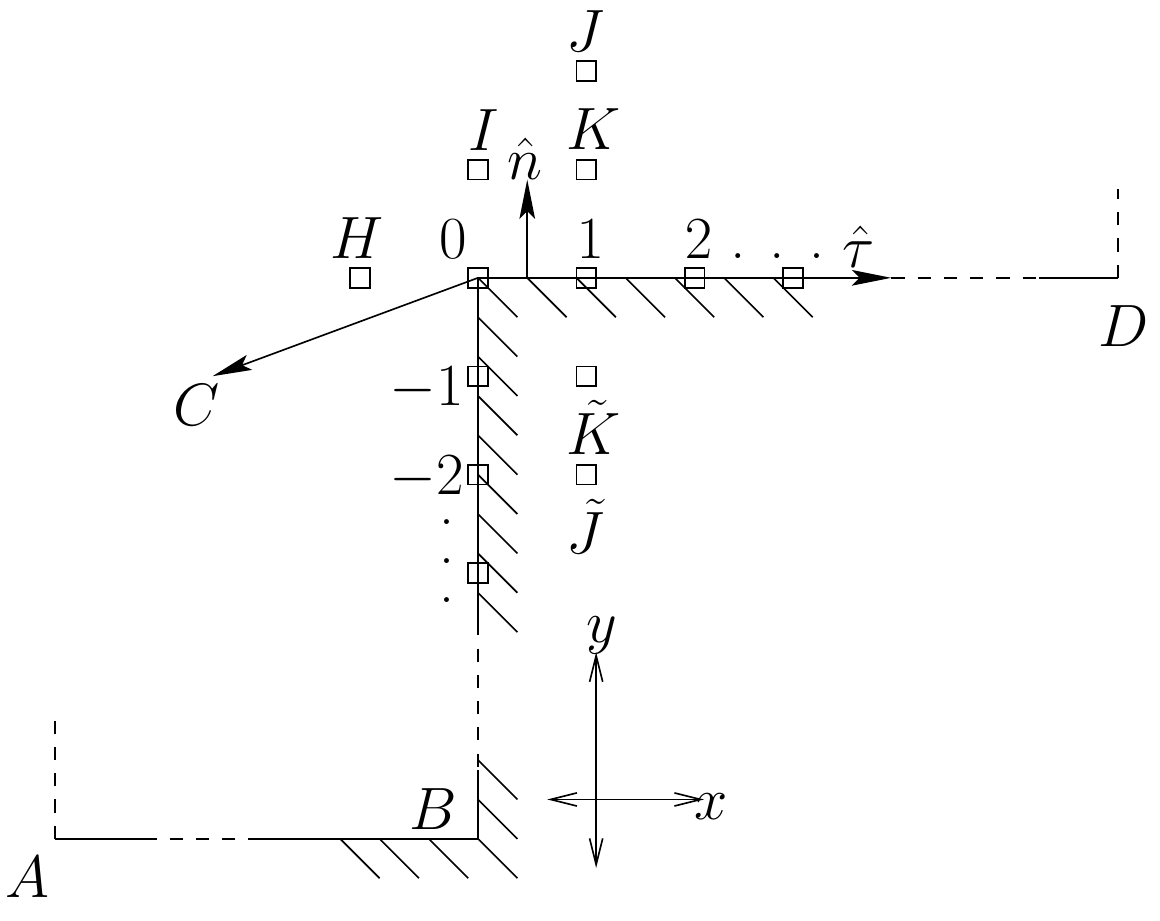}
  \caption{Mesh with grid point at corner. Interior grid points $H$, $I$, $J$ and $K$ are shown along with corresponding ghost points for $J$ and $K$.}
  \label{fig:cornerSymmConds}
\end{center}
\end{figure}
The analysis is performed using the grid points shown in figure~\ref{fig:cornerSymmConds}. On the wall, the tangent and normal are defined except at grid point $0$.
The limit of the equations for the normal component of momentum equation approaching the wall, for portions of the wall $CD$ and $BC$ are given by
\begin{equation}
  \label{eq:yMomequation}
  \frac{\partial }{\partial t}(\rho v) + \frac{\partial }{\partial y}(\rho v^2 + p) + \frac{\partial }{\partial x}(\rho u v) = 0 \text{ (for CD)},
\end{equation}
\begin{equation}
  \label{eq:xMomequation}
  \frac{\partial }{\partial t}(\rho u) + \frac{\partial }{\partial x}(\rho u^2 + p) + \frac{\partial }{\partial y}(\rho u v) = 0 \text{ (for BC)},
\end{equation}
respectively. For brevity, equations~\ref{eq:yMomequation} and \ref{eq:xMomequation} will be represented using the following equation
\begin{equation}
  \label{eq:normMomentumEquation}
  \frac{\partial }{\partial t}(\rho V_n) + \frac{\partial }{\partial n}(\rho V_n^2 + p) + \frac{\partial }{\partial \tau}(\rho V_n V_{\tau}) = 0,
\end{equation}
where $n$ (see figure~\ref{fig:cornerSymmConds}) is measured in the direction normal to the wall, $\tau$ measured along the wall.
Therefore while calculating approximation to $\tau$ derivatives at grid points on the line through $C$ and $D$, $(V_n, V_{\tau}) = (v, u)$.
For $\tau$ derivatives at grid points on the line through $B$ and $C$, $(V_n, V_{\tau}) = (u, v)$.

Analytically, on portions of the wall BC and CD (see figure~\ref{fig:cornerSymmConds}), it can be asserted that the term 
$\partial (\rho V_n V_{\tau})/\partial \tau$ is zero at every point except the corner C (grid point $0$).
Also, as mentioned earlier, determining the state at $0$ is problematic. 
In order to tackle this `corner point state problem', we used the `corner fix'
(see section~\ref{Sec:tacklingCornerPt}) of solving the discretised governing equations at $0$,  as is done at any interior grid point.

Let the velocity vector at grid point $0$ be $\vec{V}_0$ and the normal at grid point $1$ be $\hat{n}_1$. Now, $\vec{V}_0.\hat{n}_1$ need not be zero, for a $\vec{V}_0$
obtained using the `corner fix'.
In other words, velocity at grid point $0$ need not be in the same direction as the tangent at $1$. Unfortunately, calculating an upwind biased approximation to 
$\partial (\rho V_n V_{\tau})/\partial \tau$ at grid point $1$,
requires that the state at grid point $0$ also be used. Approximation to
$\partial (\rho V_n V_{\tau})/\partial \tau$, calculated using state at $0$ will not be zero if $\vec{V}_0.\hat{n}_1$ is not equal to zero, which happens to
be the case for the $\vec{V}_0$ obtained using the `corner fix'. Even if a different fix is used and a state at grid point
$0$ is assigned such that $\vec{V}_0.\hat{n}_1=0$, it will lead to a similar problem at grid point $-1$ unless $\vec{V}_0$ is set equal to $\vec{0}$. But
grid point $0$ is not a stagnation point. Therefore whatever non-zero velocity is assigned at grid point $0$, it will lead to a non-zero 
$\partial (\rho V_n V_{\tau})/\partial \tau$ at either grid point $1$ or $-1$ or at both of them.

Using SST at grid point $1$ will lead to $\partial(\rho V_n^2 + p)/\partial n = 0$.
$\partial (\rho V_n V_{\tau})/\partial \tau \neq 0$ and $\partial(\rho V_n^2 + p)/\partial n = 0$  will
lead to a nonzero normal component of momentum at grid point $1$ through equation~(\ref{eq:normMomentumEquation}).
Repeating this argument, this non zero normal component of velocity at grid point $1$ will result in a non zero normal component of velocity at grid point $2$. In 
subsequent time steps,
this nonzero normal velocity will propagate to other grid points along the wall. The same will happen at grid points $-1$, $-2$, ..., which are below grid point $0$.
Different ways to address this flow leak for these problems are described next.

%This kind of a flow leak will also occur for geometries with corners such as the ones shown in figure \ref{fig:stepWithOutDownstreamOfCorner}. 
%As indicated in the figure, a portion of
%the boundary, BC, is also a wall. A problem similar to that at grid points $1$ and $-1$ will occur at point C due to using SST. This will lead to flow leak along BC.
%\begin{figure}[!htbp]
%\begin{center}
%  \includegraphics[width=0.4\textwidth]{images/stepWithOutDownstreamOfCorner-eps-converted-to.pdf}
%  \caption{Sketch of a flow domain with a different type of corner ($C$). AB, BC and DE are walls.}
%  \label{fig:stepWithOutDownstreamOfCorner}
%\end{center}
%\end{figure}

\subsection{Algorithmic fixes for the flow leak problem}\label{Sec:AlgoFixes}
If no-penetration condition is satisfied
at grid points $-1$ and $1$, the nonlinear WENO
weights will make sure that $ \partial (\rho V_n V_{\tau})/\partial \tau$ is essentially zero at grid points below $-1$ and grid points to the right of $1$. This, along
with SST will prevent mass leak along the wall at all grid points except the corner. Two algorithmic fixes to achieve this are given next. 

The first fix is to set the normal component of velocity to zero (enforcing no-penetration), at grid points $-1$ and $1$, after each time step or Runge-Kutta stage.
This, as mentioned earlier, will prevent the flow leak from happening at grid points below $-1$ and grid points to the right of $1$.
This technique will be referred to as `SSTNPE' (Standard symmetry technique with no penetration enforced).

\begin{figure}[!htbp]
  \centering
  \begin{subfigure}{.50\textwidth}
    \centering
  \includegraphics[width=0.6\textwidth]{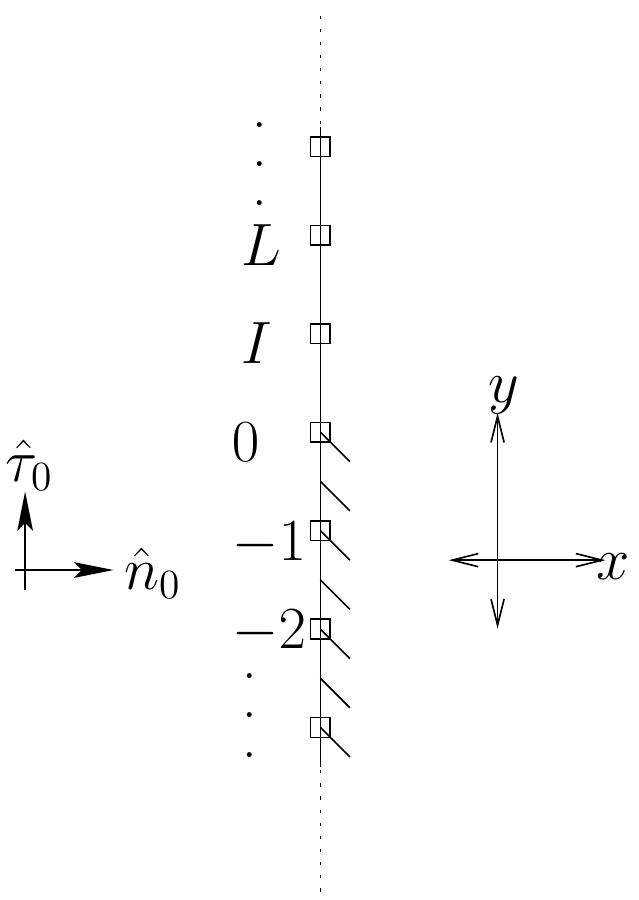}
  \caption{Along grid line parallel to $y$-axis}
  \label{fig:gridLineParallelToYAxis}
\end{subfigure}%
\begin{subfigure}{.39\textwidth}
  \centering
  \includegraphics[width=0.98\textwidth]{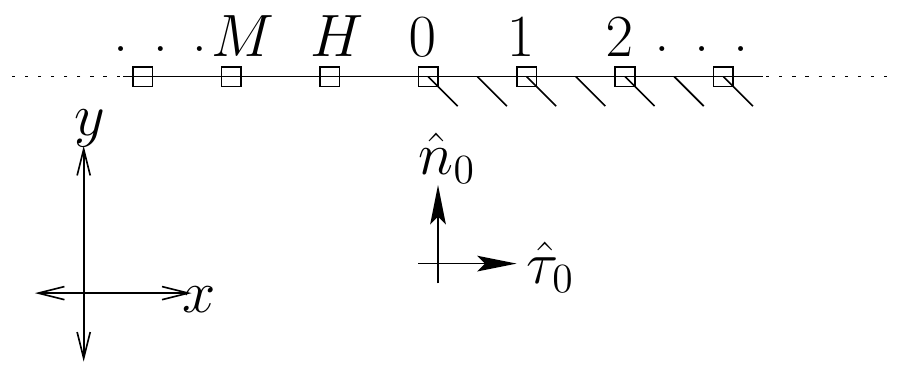}
  \caption{Along grid line parallel to $x$-axis}
  \label{fig:gridLineParallelToXAxis}
\end{subfigure}
\caption{$\hat{\tau}_0$ and $\hat{n}_0$ for the SSTCVD fix.}
  \label{fig:gridLinesThruCorner}
\end{figure}
The second fix is called the corner velocity direction fix. Let the density, momentum density, and total energy density at the grid point $0$, obtained using the 
`corner fix' be $\rho_0,\, \rho_0 \vec{V}_0,\, \rho_0 e_{t_{0}}$, respectively. Let $|\vec{V}_0| = V_0$. Let $\hat{e}_x$ and $\hat{e}_y$ be the unit vectors along the positive
$x$ and $y$
directions, respectively. There are two grid lines through grid point $0$ as shown
in figure~\ref{fig:gridLinesThruCorner}, one parallel to $x$-axis (figure~\ref{fig:gridLineParallelToXAxis}) and one parallel to $y$-axis 
(figure~\ref{fig:gridLineParallelToYAxis}). For calculating $x$ and $y$ derivatives on these grid lines, except at grid point $0$,
the following state at grid point $0$ is used: $\rho_0,\, \rho_0 V_0\hat{\tau_0},\, \rho_0 e_{t_{0}}$. As shown in the
figure~\ref{fig:gridLinesThruCorner}, $\hat{\tau_0} = \hat{e}_y$
for grid points along the grid line parallel to $y$-axis and $\hat{\tau_0} = \hat{e}_x$ for grid points along the grid line parallel to $x$-axis.
That is, while calculating $x$ derivatives  
(like $ \partial(\rho u^2 +p)/\partial x $ and
$ \partial (\rho u v)/\partial x$) at grid points like  $H$ and $1$ (figure~\ref{fig:cornerSymmConds}), the velocity
at grid point $0$ is taken to be $V_0\hat{e}_x$. Similarly, while calculating $y$ derivatives (like $ \partial(\rho v^2 +p)/\partial y $ and 
$ \partial (\rho u v)/\partial y$) at grid points like $I$ and $-1$, 
the velocity at grid point $0$ is taken to be $V_0\hat{e}_y$.
%Therefore, for grid points along CD
%(see figure \ref{fig:cornerSymmConds}), like $H$, $1$ ...,  a state at grid point $0$ with
%velocity along the positive $x$ direction is used. For grid points along BC, like $I$, $-1$, ...,  a state at grid point $0$ with velocity along the positive $y$ direction
%will be used.
The WENO weights and prescribed direction of 
velocity will ensure $ \partial (\rho V_n V_{\tau})/\partial \tau$ will be zero at grid points $-1$ and $1$, which along with SST will ensure no-penetration
condition is satisfied at grid points $-1$ and $1$. This fix will be referred to as `SSTCVD' (Standard symmetry technique with corner velocity direction fix).
This is similar to the suggestion of Verhoff \cite{verhoff2004} that ``At the corner points the velocity (or momentum) vector rotates at constant magnitude through an angle
$\alpha$ [$\alpha$ is the \emph{flow turning angle} which is equal to $\pi/2$ for problems considered in this paper] due to an impulsive-type interaction''.
%It must be noted that the SSTCVD fix to mitigate flow leak cannot be used for corners such as the point C shown in figure \ref{fig:stepWithOutDownstreamOfCorner}

Using either `SSTNPE' or `SSTCVD', the no-penetration condition will be satisfied at 
all grid points on the wall except the
corner. However, it will
lead to contribution of the non-zero term `$ \partial (\rho V_n V_{\tau})/\partial \tau$', at grid points $1$ and $-1$, being ignored and a corresponding error in the state at
grid points $1$ and $-1$.
To avoid this,
we retain this non-zero derivative and propose a new technique to apply wall boundary
conditions at grid points $1$ and $-1$. In the next section, we propose modifications to the standard symmetry technique so that the effect of the non-zero term,
$\partial (\rho V_n V_{\tau})/\partial \tau$, is also considered.

\section{Modified symmetry technique for walls with expansion corners}\label{Sec:MST}

Now, we propose modifications to the standard symmetry technique to incorporate the non-zero tangential derivative, $ \partial (\rho V_n V_{\tau})/\partial \tau$, near
corners and derive equations to be used at grid points near corners for applying boundary conditions.

\subsection{Condition on normal derivative of pressure on the wall, near corners}
The governing differential equations are solved at the corner grid point (point 0 in figure~\ref{fig:cornerSymmConds}). 
The equations that will be derived next, are for applying boundary conditions at grid points adjacent to the corner, that is for grid points $-1$ and $1$. 
At grid points near the corner, the normal direction is defined. The limit of the normal momentum equation approaching the wall is given by equation~(\ref{eq:normMomentumEquation}) and is 
repeated below.
\begin{equation*}%\label{eq:normMomEqn}
  \frac{\partial }{\partial t}(\rho V_n) +  \frac{\partial}{\partial n} (\rho V_n^2 +p) + \frac{\partial }{\partial \tau}(\rho V_n V_{\tau}) = 0.
\end{equation*}
Near the corner, on the wall, no penetration implies
$\rho V_n =0$ and $\partial (\rho V_n)/\partial t=0$. That leaves us with an equation

\begin{equation}\label{eq:normPresGrad}
  \frac{\partial }{\partial n}(\rho V^2_n) + \frac{\partial p}{\partial n} + \frac{\partial }{\partial \tau}(\rho V_n V_{\tau}) = 0, \text{\emph{ near} corners}
\end{equation}
To ensure that the normal component of velocity on the wall is zero, the states in the ghost points must be such that they satisfy the discretised version of equation
(\ref{eq:normPresGrad}).
It is pointed out that while discretising equation~(\ref{eq:normPresGrad}), the derivatives should be calculated using the Lax-Freidrichs flux splitting. 
 
For a third order scheme, there will be 2 layers of ghost points (as shown in figure~\ref{fig:cornerSymmConds}) and values for density, pressure, and velocity are needed at these
points in these 2 layers. For ensuring free slip we use the equations in the standard symmetry technique written in section~\ref{Sec:SymmCondAndProbs} except for pressure and density. As mentioned earlier,
discretised form of equation~(\ref{eq:normPresGrad}) is used to calculate pressure gradient. We also need the gradient of density (or temperature) to define the states at the ghost points. 
%A zero temperature gradient normal to wall is the condition usually prescribed for an adiabatic wall when heat flux is also modeled in the governing equations. Though heat flux
%is
%not modeled in the compressible Euler Equations, dissipation due to flux splitting in the energy equation is similar to the dissipation due to heat flux.
A zero normal temperature gradient is chosen for applying the boundary conditions.
To calculate density gradient normal to the wall the equation of state $p=\rho R T$, with $R=1$ is used.

Taking a derivative of the equation of state $p=\rho T$ along $\hat{n}$, we get
\begin{equation}\label{eq:eqStateNormGrad}
  \frac{\partial p}{\partial n} = \frac{\partial \rho}{\partial n}  T + \rho \frac{\partial T}{\partial n}
\end{equation}
Setting gradient of temperature normal to wall in equation~(\ref{eq:eqStateNormGrad}) to zero, we
get the following:
\begin{equation}\label{eq:normDenGrad}
  \frac{\partial p}{\partial n} = \frac{\partial \rho}{\partial n} T
\end{equation}

Using the discrete form of equations~(\ref{eq:normPresGrad}) and (\ref{eq:normDenGrad}) at $-1$, 
we must determine pressure and density at grid points at $\tilde{J}$ and $\tilde{K}$. This is not possible as there are only 2 equations but 4 unknowns. To avoid this problem
we can set pressure and density  at one of the points $\tilde{J}$ and $\tilde{K}$ equal to that of $J$ and $K$, respectively, that is $p_{\tilde{J}} = p_J, \rho_{\tilde{J}} = \rho_J$
or, $p_{\tilde{K}} = p_K, \rho_{\tilde{K}} = \rho_K$.
Choosing $p_{\tilde{J}} = p_J, \rho_{\tilde{J}} = \rho_J$ is better because depending on the WENO weights, the normal pressure derivative on the wall may be independent of
pressure and density at grid point $\tilde{J}$.
%explain that choosing $p_{K'} = p_K$ is better because one gets a 7th degree polynomial in PL' which is guaranteed to have at least one real root
Therefore, we choose the equation $p_{\tilde{J}} = p_J, \rho_{\tilde{J}} = \rho_J$ to eliminate the pressure and density at $\tilde{J}$ which leaves us with two variables and
two equations which can be solved. Using discrete forms of equations~(\ref{eq:normPresGrad}) and (\ref{eq:normDenGrad}), we solve for pressure and density at the grid 
point $\tilde{K}$.
A linear central difference with formal order of accuracy of 4, was used to
discretise equation~(\ref{eq:normDenGrad})
and the Shu-Osher conservative finite difference method with WENO-NP3 reconstruction and Lax-Freidrichs flux splitting was used to discretise equation~(\ref{eq:normPresGrad}). For
WENO-NP3 reconstruction, equation~(\ref{eq:normPresGrad}), is non-linear in $\rho_{\tilde{K}}$ and $p_{\tilde{K}}$.
Bisection method was used to solve discretized forms of equations~(\ref{eq:normPresGrad}) and (\ref{eq:normDenGrad}) to an accuracy of $10^{-8}$.
This \textbf{m}odified \textbf{s}ymmetry \textbf{t}echnique described above will be referred to as MST.
\subsection{Algorithm for solving for pressure in MST using bisection method \cite{conte2017} }
\begin{enumerate}
  \item Choose an initial bounding interval for $p_{\tilde{K}}$: $[a, b]\text{, where }a<b$\label{step:firstStep}
  \item Calculate densities at ghost point $\tilde{K}$ ($\rho_{\tilde{K}}$) using equation~(\ref{eq:normDenGrad}) with $p_{\tilde{K}}=a$ and $p_{\tilde{K}}=b$ and let them be 
    $\rho^a_{\tilde{K}}$, $\rho^b_{\tilde{K}}$, respectively.
  \item Check if setting $p_{\tilde{K}}=a, \rho_{\tilde{K}} = \rho^a_{\tilde{K}}$ produces numerical approximation to LHS of equation~(\ref{eq:normPresGrad}) with opposite
    sign to that obtained by setting  $p_{\tilde{K}}=b\text{ and } \rho_{\tilde{K}} = \rho^b_{\tilde{K}}$. Else, go back to step~\ref{step:firstStep} and change the bounding
    interval. \label{step:sanityCheckCTT}
  \item Using the bounding interval of pressure ($[a, b]$) and using a 4\textsuperscript{th} order accurate  discretisation of equation~(\ref{eq:normDenGrad}), calculate three
  densities at $\tilde{K}$ corresponding to $p_{\tilde{K}}=a$, $p_{\tilde{K}}=(a+b)/2$ and $p_{\tilde{K}}=b$ and let those densities be $\rho^a_{\tilde{K}},
  \rho^{(a+b)/2}_{\tilde{K}}, \rho^b_{\tilde{K}}$, respectively.\label{step:begCTT}
  \item Calculate three numerical approximations to the LHS in equation~(\ref{eq:normPresGrad}) using flux splitting according to the procedure given in
    sections~\ref{Sec:shuOsherConsFinDiffSch} - \ref{Sec:WENOJS3}, by using ${(p_{\tilde{K}}, \rho_{\tilde{K}}) = (a, \rho^a_{\tilde{K}})}$,
    ${(p_{\tilde{K}}, \rho_{\tilde{K}}) = ((a+b)/2, \rho^{(a+b)/2}_{\tilde{K}})}$, ${(p_{\tilde{K}}, \rho_{\tilde{K}}) = (b, \rho^b_{\tilde{K}})}$ and label them $R_1, R_2, R_3$, respectively.
  \item Using $R_1, R_2\text{, and } R_3$, choose a new, smaller bounding interval for $p_{\tilde{K}}$.  If $(R_1R_2)<0$ the new interval is
  $[a, (a+b)/2]$, otherwise it is $[(a+b)/2, b]$.
  \item With the new interval obtained in the above step, repeat from step~\ref{step:begCTT} till  values of pressure and density are obtained such that
    equation~(\ref{eq:normPresGrad}) is satisfied with an error of $10^{-8}$.
  \end{enumerate}

 The initial bounding interval for pressure for starting the bisection method is chosen as the interval $[-\eta p_{prev}, \eta p_{prev}]$, where $\eta>0$ and
 $p_{prev}$ is the root of equation~(\ref{eq:normPresGrad}) in previous time step or RK Stage and $\eta$ is chosen so as to satisfy the condition in step~\ref{step:sanityCheckCTT}.

 \subsection{A note on flow field initialisation and boundary conditions and CFL number}\label{Sec:InitPrecautions}
For ease of initialisation, the flow field may be initialised with uniform flow at all grid points including the grid points on the wall. This will lead to the normal component
of velocity not being zero at some grid points on the wall, initially. In such a case, in addition to applying MST, the normal component of
velocity must be set to zero after every RK stage for a few thousand time steps until the normal component of momentum on the wall settles to zero or a very low value. Also,
for the first few thousand time steps, a CFL number of $1/128$ should be used and later (after 5000 or 10000 time steps) it can be increased to a higher value like $1/8$ or 
$1/4$. Failing to do this may lead to severe convergence problems.

%***********Additions
%The modified symmetry technique can also be used for applying wall boundary conditions for walls with the type of corners shown in figure
%\ref{fig:stepWithOutDownstreamOfCorner}. But we restrict the scope of this paper to solving flow over a forward facing and backward facing steps.
The algorithmic fixes and modified symmetry technique described in previous sections are used to obtain numerical solution to flows over backward and 
forward facing step. A comparison of
solutions obtained using these different techniques is presented in the next section.

\section{Testing the new boundary technique (MST) and algorithmic fixes}\label{Sec:BCsTesting}
The modified symmetry technique (MST) described in section~\ref{Sec:MST} and the algorithmic fixes described in section~\ref{Sec:AlgoFixes} are tested by solving supersonic 
flows over
forward facing and backward facing step. A comparison of numerical solutions obtained using these different techniques for meshes with different grid point spacings is 
presented. Labels for the
five different wall boundary condition techniques (WBCTs) are given below:
\begin{enumerate}[label=(\alph*)]
  \item Standard symmetry technique with \textbf{n}o \textbf{g}rid points on the \textbf{w}all and at corner - SSTNGW,
  \item Standard symmetry technique with \textbf{g}rid points on the \textbf{w}all and corner, and the corner fix being used - SSTGW,
  \item Standard symmetry technique with grid points on the wall, with the corner fix being used, and \textbf{n}o \textbf{p}enetration
    \textbf{e}nforced near the corner - SSTNPE (refer to section~\ref{Sec:AlgoFixes}),
  \item Standard symmetry technique with grid points on the wall, with corner fix being used, and modification of corner velocity direction  - SSTCVD (refer to
    section~\ref{Sec:AlgoFixes}),
  \item \textbf{M}odified \textbf{s}ymmetry \textbf{t}echnique - MST (refer to section~\ref{Sec:MST}).
  \end{enumerate}

We start with the flow over forward facing step.

\subsection{Mach 4.0 flow over a forward facing step}
The problem domain and boundary conditions were described in section~\ref{Sec:FowFacStepGeo}.
%The lengths of different portions of the flow field are, AB = $0.6$ units,
%CD = $0.6$ units,
%AG = $1.0$ units, BC = $0.2$ units, CF = $0.8$ units.
%The flow field is initialized with $M = 4.0$ with the velocity vector in x direction, $\rho = 1.4$, $p = 1.0$, which makes speed of sound $a=1$ and $\alpha =5.0$($=1.0+M$).

\begin{figure}[!htbp]
  \begin{subfigure}{.49\textwidth}
    \centering
    \includegraphics[width=.9\textwidth]{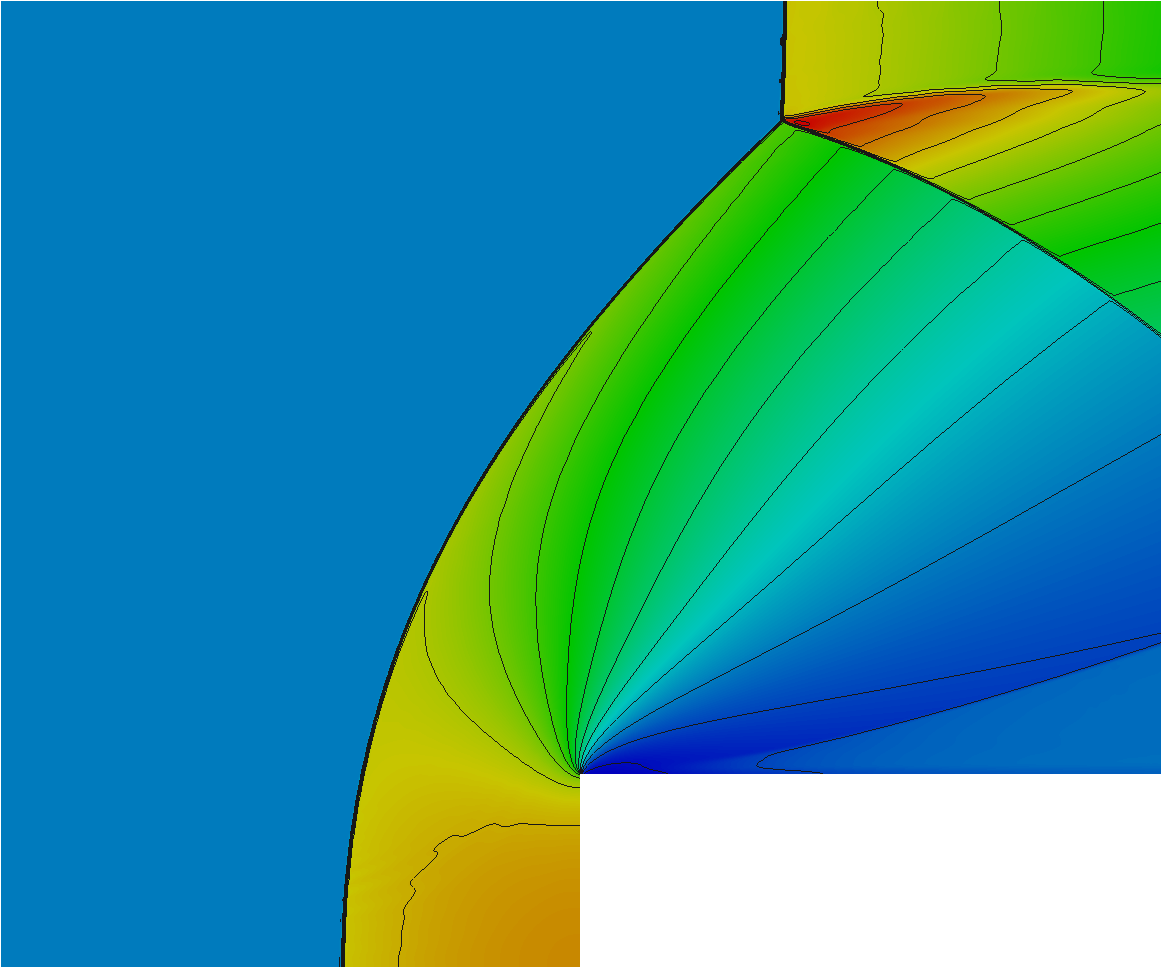}
    \caption{SSTNGW}
    \label{fig:M40DX1By800SSTNGW}
  \end{subfigure}%
  \begin{subfigure}{.49\textwidth}
    \centering
    \includegraphics[width=.9\textwidth]{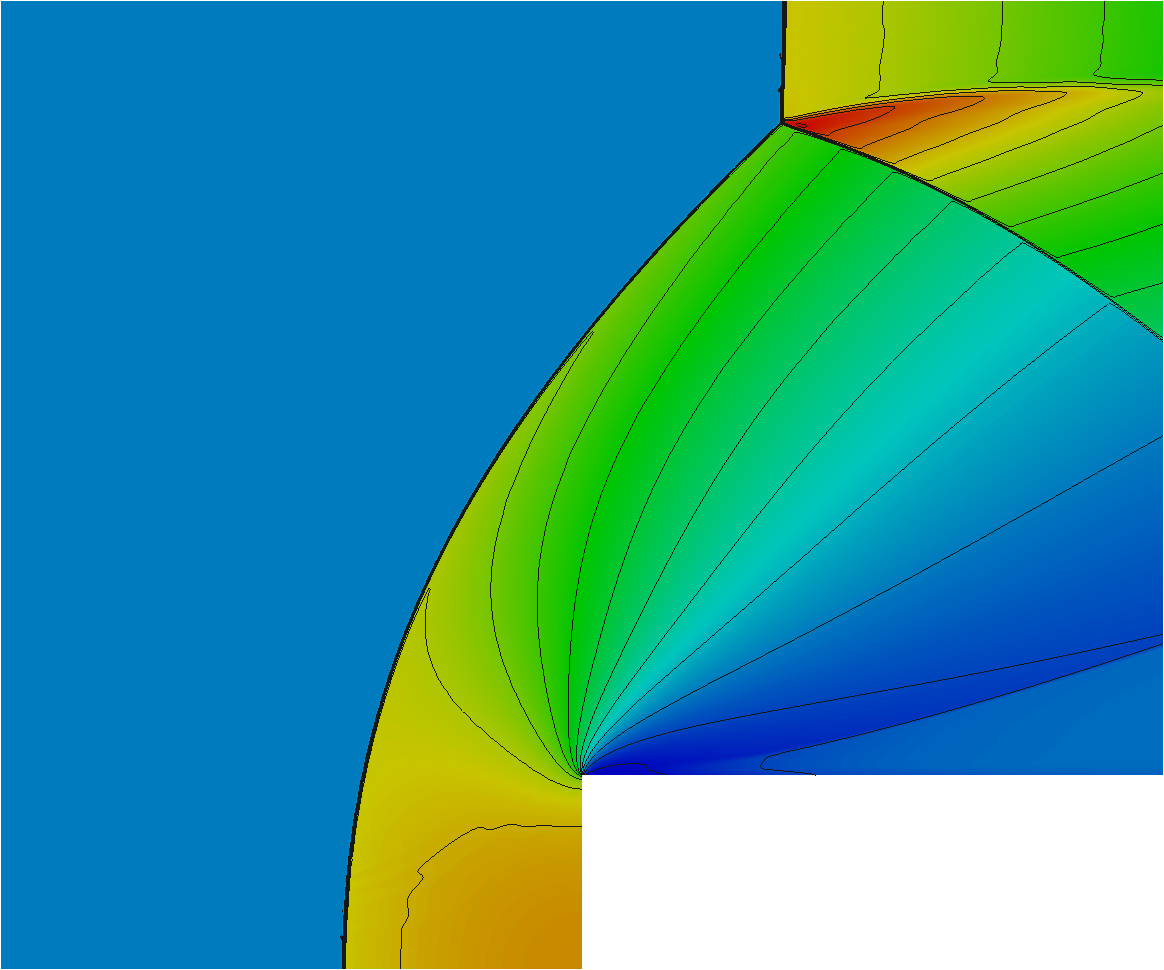}
    \caption{SSTGW}
    \label{fig:M40DX1By800SSTGW}
  \end{subfigure}
  \begin{subfigure}{.49\textwidth}
    \centering
  \includegraphics[width=.9\textwidth]{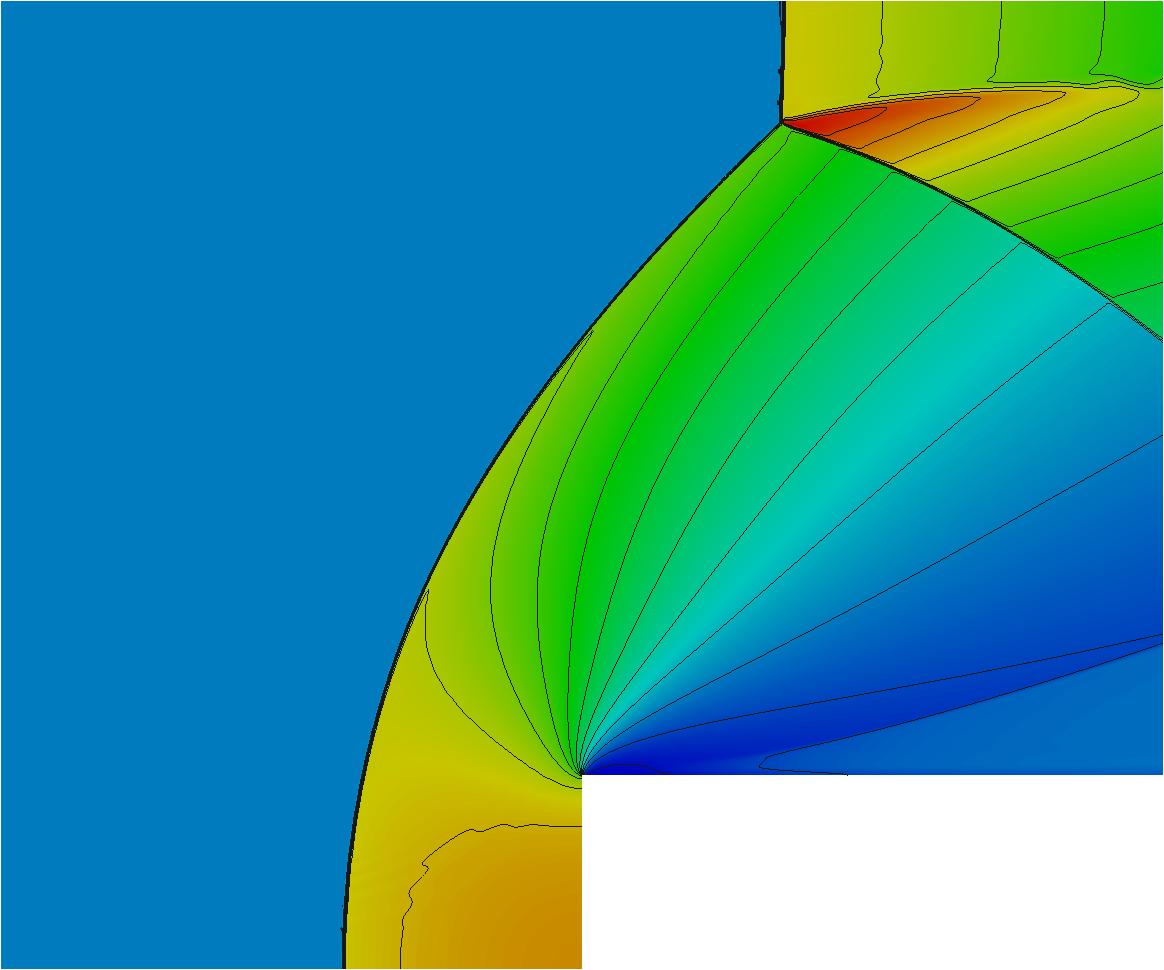}
  \caption{SSTNPE}
  \label{fig:M40DX1By800SSTNPE}
\end{subfigure}%
\begin{subfigure}{.49\textwidth}
  \centering
  \includegraphics[width=.9\textwidth]{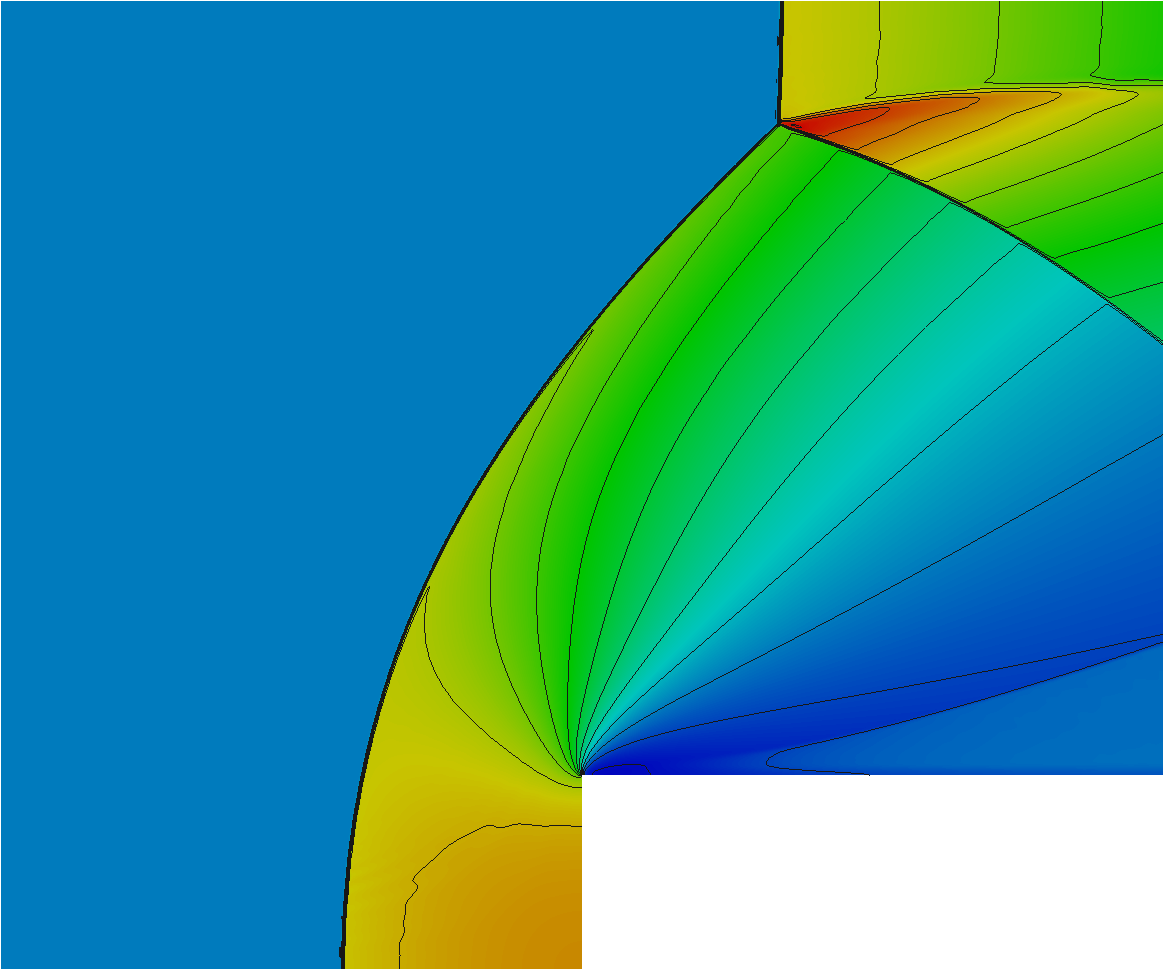}
  \caption{SSTCVD}
  \label{fig:M40DX1By800SSTCVD}
\end{subfigure}
\begin{subfigure}{.49\textwidth}
  \centering
  \includegraphics[width=.9\textwidth]{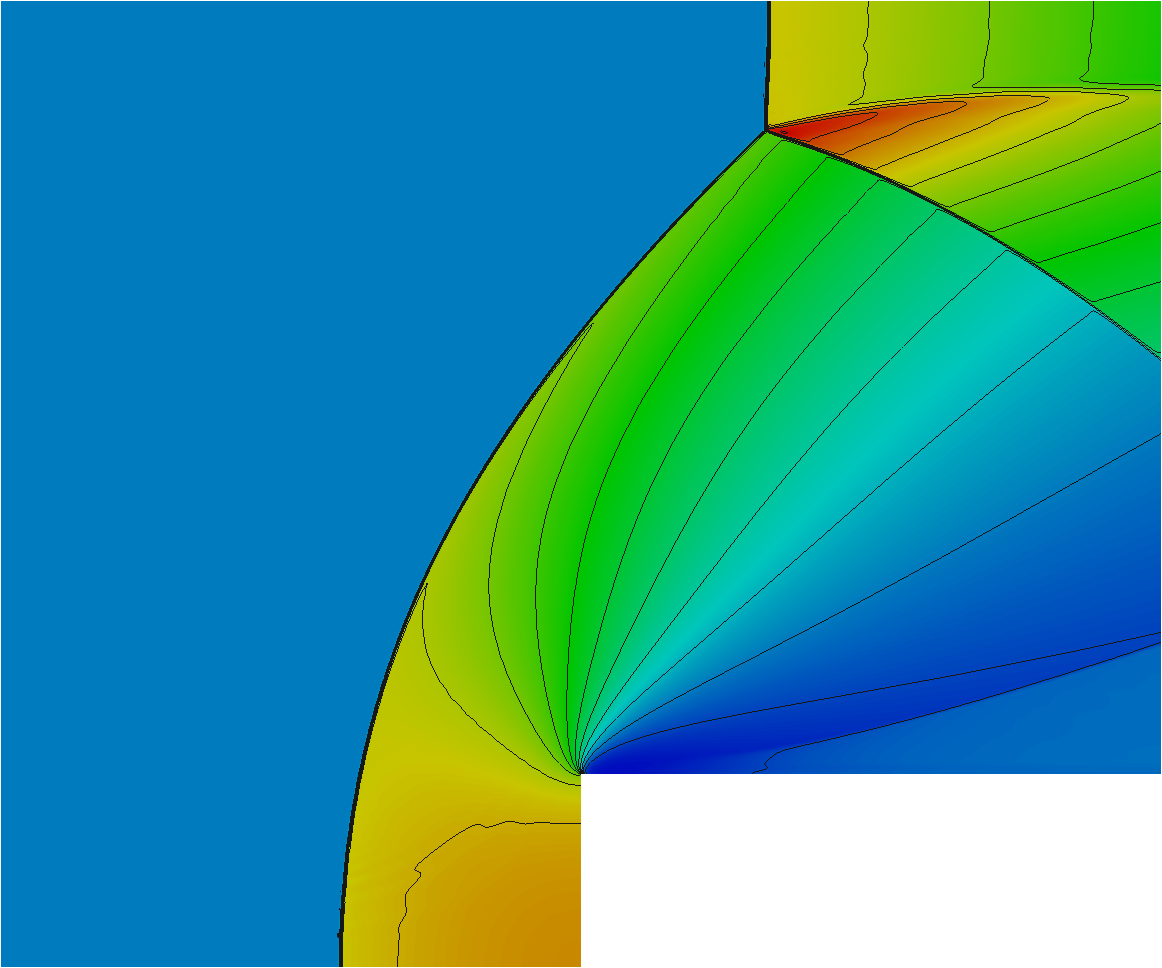}
  \caption{MST}
  \label{fig:M40DX1By800MST}
\end{subfigure}%
\begin{subfigure}{.49\textwidth}
  \centering
  \includegraphics[width=.3\textwidth]{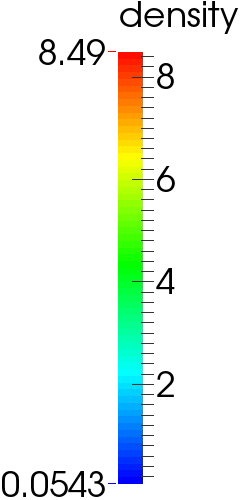}
  \caption{Color Map}
  \label{fig:M40DX1By800ColMap}
\end{subfigure}
\caption{Color plot of density with 15 equally spaced density contours for Mach 4.0 flow over a forward facing step with GPS = $1/800$ and the five different WBCTs.
All techniques produce solutions with Mach reflection on the top wall GE (see figure~\ref{fig:fowFacStepGeo}), as opposed to solutions with GPS = $1/200$, for which only
MST and SSTCVD techniques produce Mach reflection.}
\label{fig:M40D10DX1By800}
\end{figure}

\begin{figure}[!htbp]
  \begin{subfigure}{.5\textwidth}
    \centering
    \includegraphics[width=.9\textwidth]{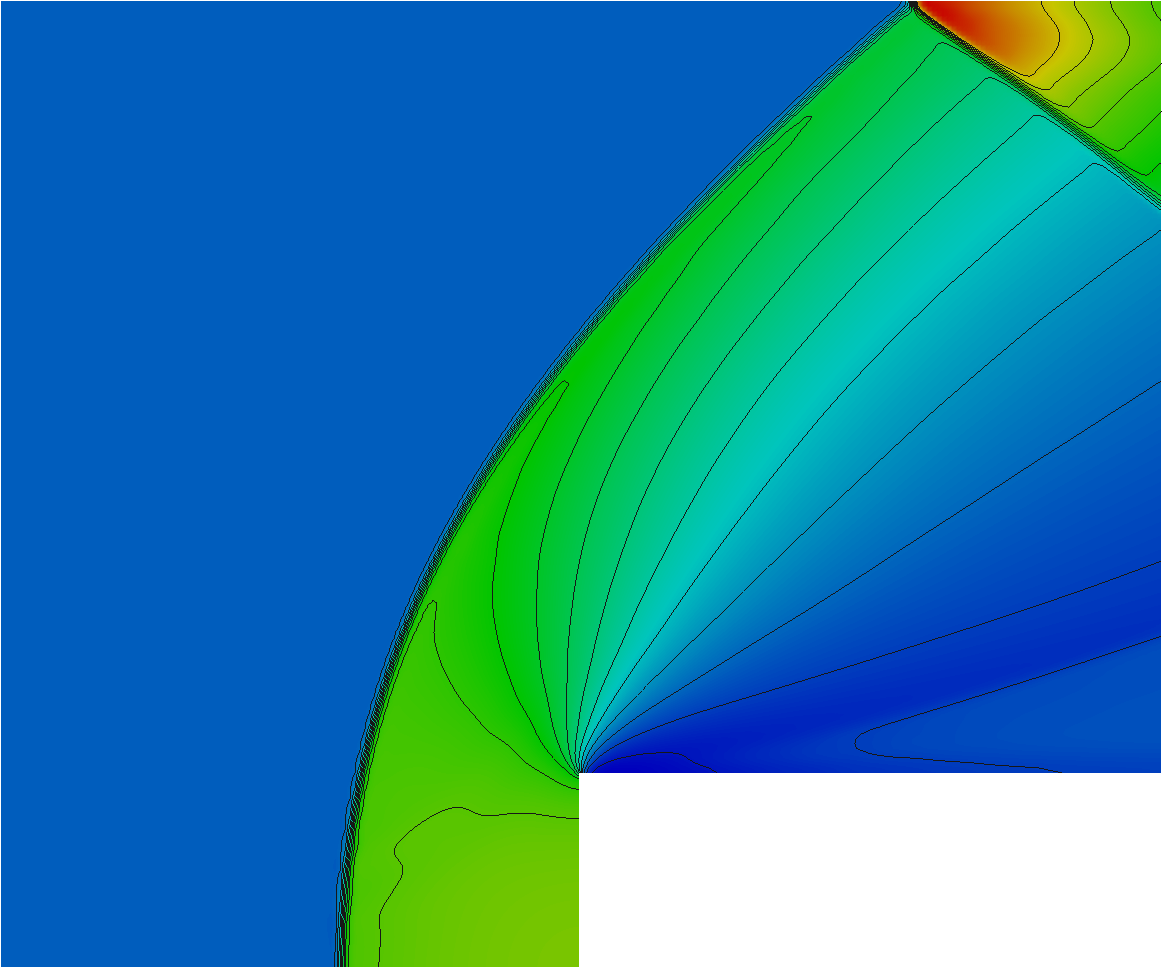}
    \caption{SSTNGW (regular reflection at top wall).}
    \label{fig:M40DX1By200SSTNGW}
  \end{subfigure}%
  \begin{subfigure}{.5\textwidth}
    \centering
    \includegraphics[width=.9\textwidth]{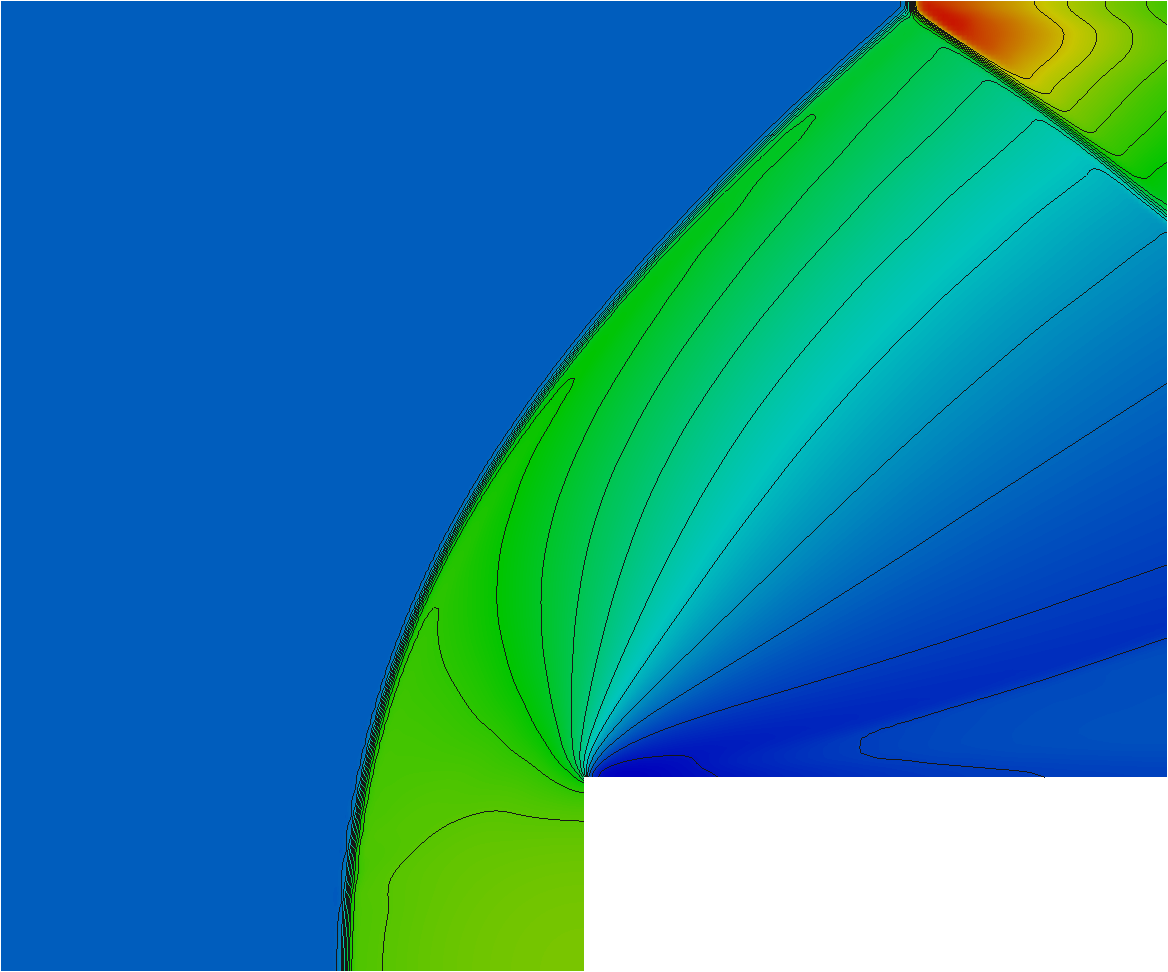}
    \caption{SSTGW (regular reflection at top wall).}
    \label{fig:M40DX1By200SSTGW}
  \end{subfigure}
  \begin{subfigure}{.5\textwidth}
    \centering
  \includegraphics[width=.9\textwidth]{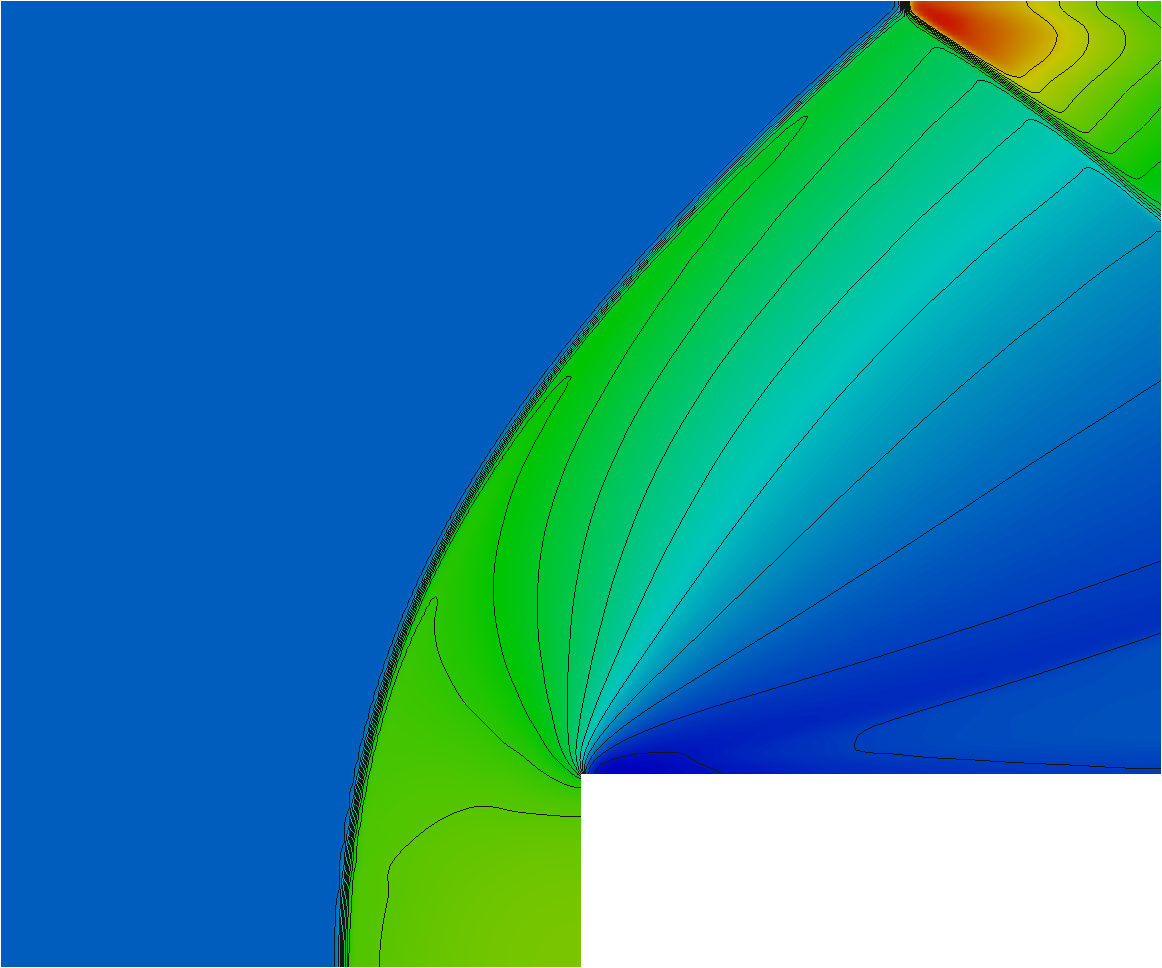}
  \caption{SSTNPE (regular reflection at top wall).}
  \label{fig:M40DX1By200SSTNPE}
\end{subfigure}%
\begin{subfigure}{.5\textwidth}
  \centering
  \includegraphics[width=.9\textwidth]{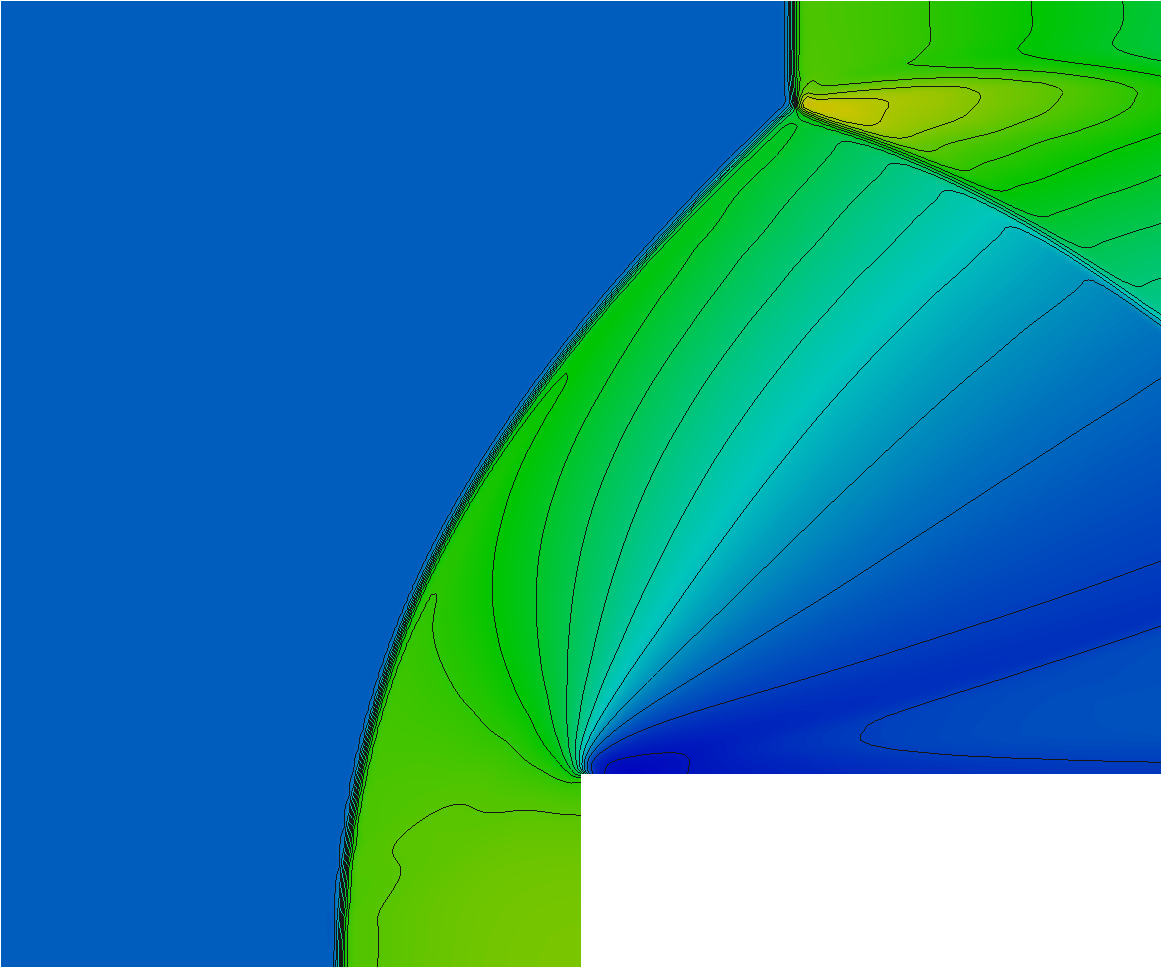}
  \caption{SSTCVD (Mach reflection at top wall).}
  \label{fig:M40DX1By200SSTCVD}
\end{subfigure}
\begin{subfigure}{.5\textwidth}
  \centering
  \includegraphics[width=.9\textwidth]{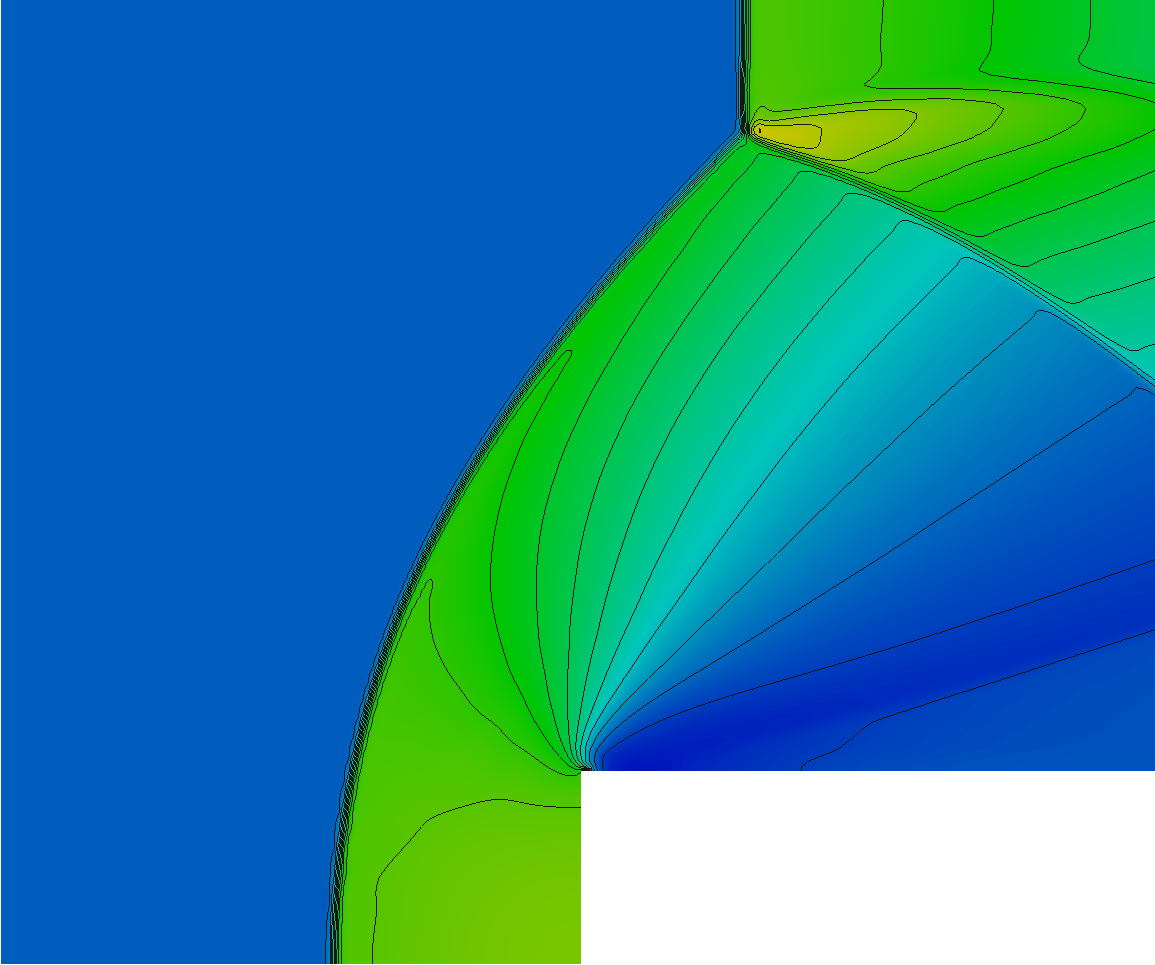}
  \caption{MST (Mach reflection at top wall).}
  \label{fig:M40DX1By200MST}
\end{subfigure}%
\begin{subfigure}{.5\textwidth}
  \centering
  \includegraphics[width=.3\textwidth]{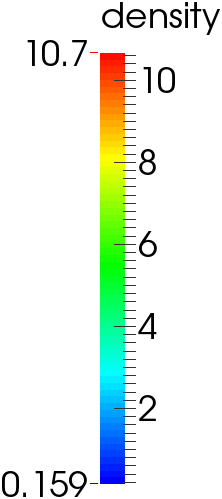}
  \caption{Color Map}
  \label{fig:M40DX1By200ColMap}
\end{subfigure}
\caption{Color plot of density with 15 equally spaced density contours for Mach 4.0 flow over a forward facing step with GPS = $1/200$ and the five different WBCTs.
All corner techniques except MST and SSTCVD fail to produce a Mach reflection on the top wall GE (see figure~\ref{fig:fowFacStepGeo})}
\label{fig:M40D10DX1By200}
\end{figure}

\begin{figure}[!htbp]
  \begin{center}
  \begin{subfigure}{.5\textwidth}
    \centering
    \includegraphics[width=.9\textwidth]{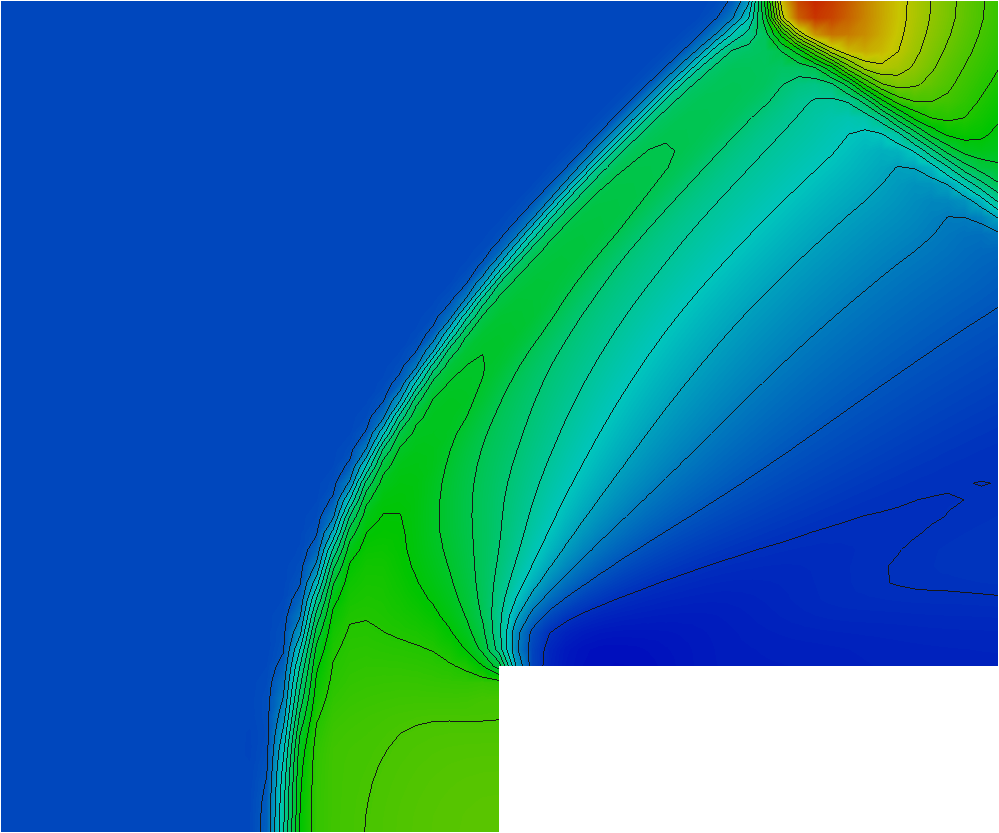}
    \caption{SSTCVD, GPS = $1/50$}
    \label{fig:SSTCVD1By50}
  \end{subfigure}%
  \begin{subfigure}{.5\textwidth}
    \centering
    \includegraphics[width=.9\textwidth]{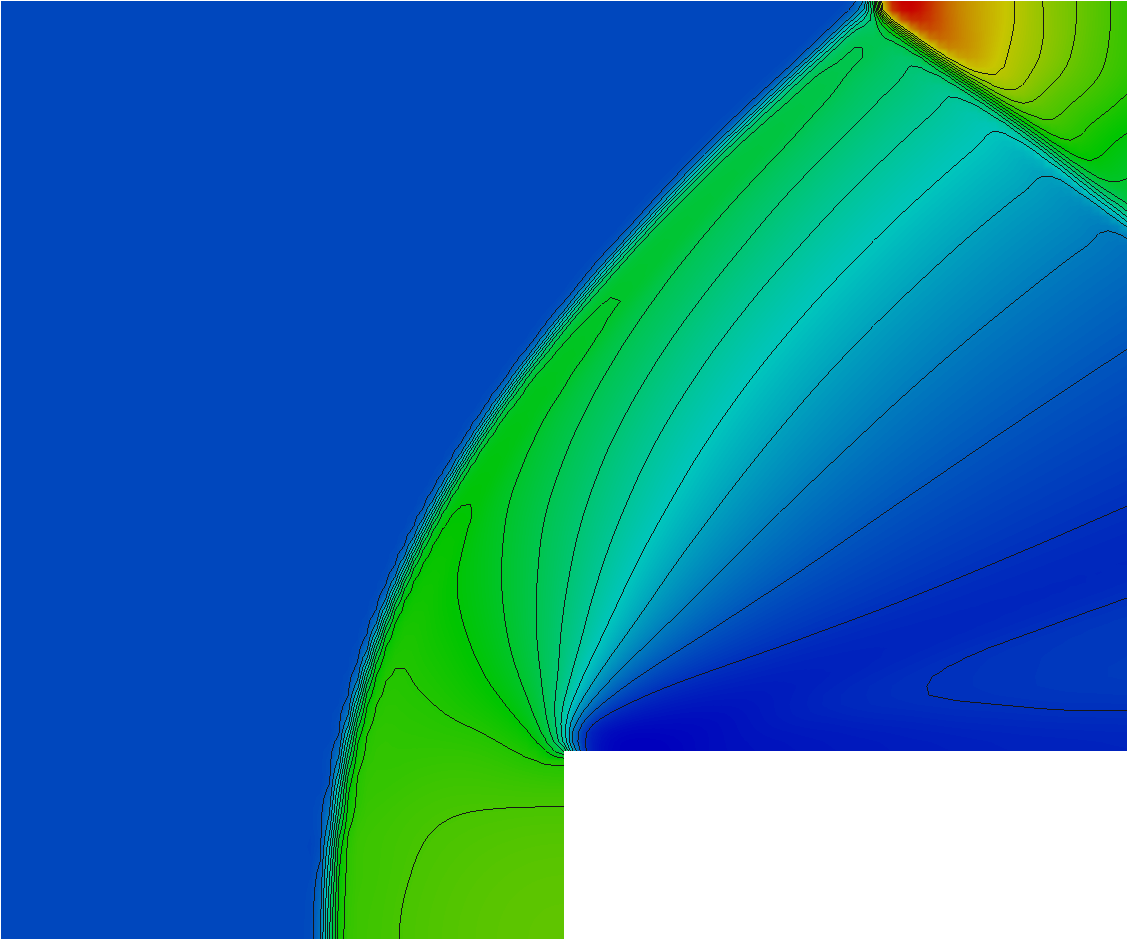}
    \caption{SSTCVD, GPS = $1/100$}
    \label{fig:SSTCVD1By100.png}
  \end{subfigure}
  \begin{subfigure}{.5\textwidth}
    \centering
    \includegraphics[width=.9\textwidth]{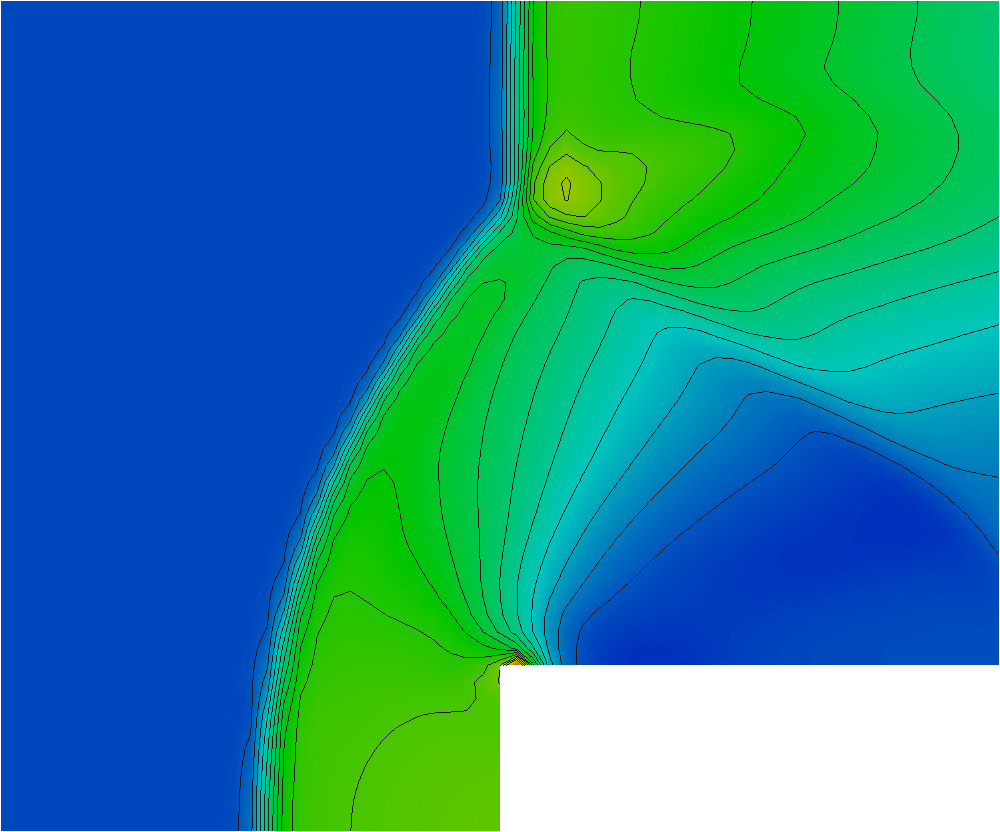}
    \caption{MST, GPS = $1/50$ (Mach reflection)}
    \label{fig:MST1By5}
  \end{subfigure}%
  \begin{subfigure}{.5\textwidth}
    \centering
    \includegraphics[width=.9\textwidth]{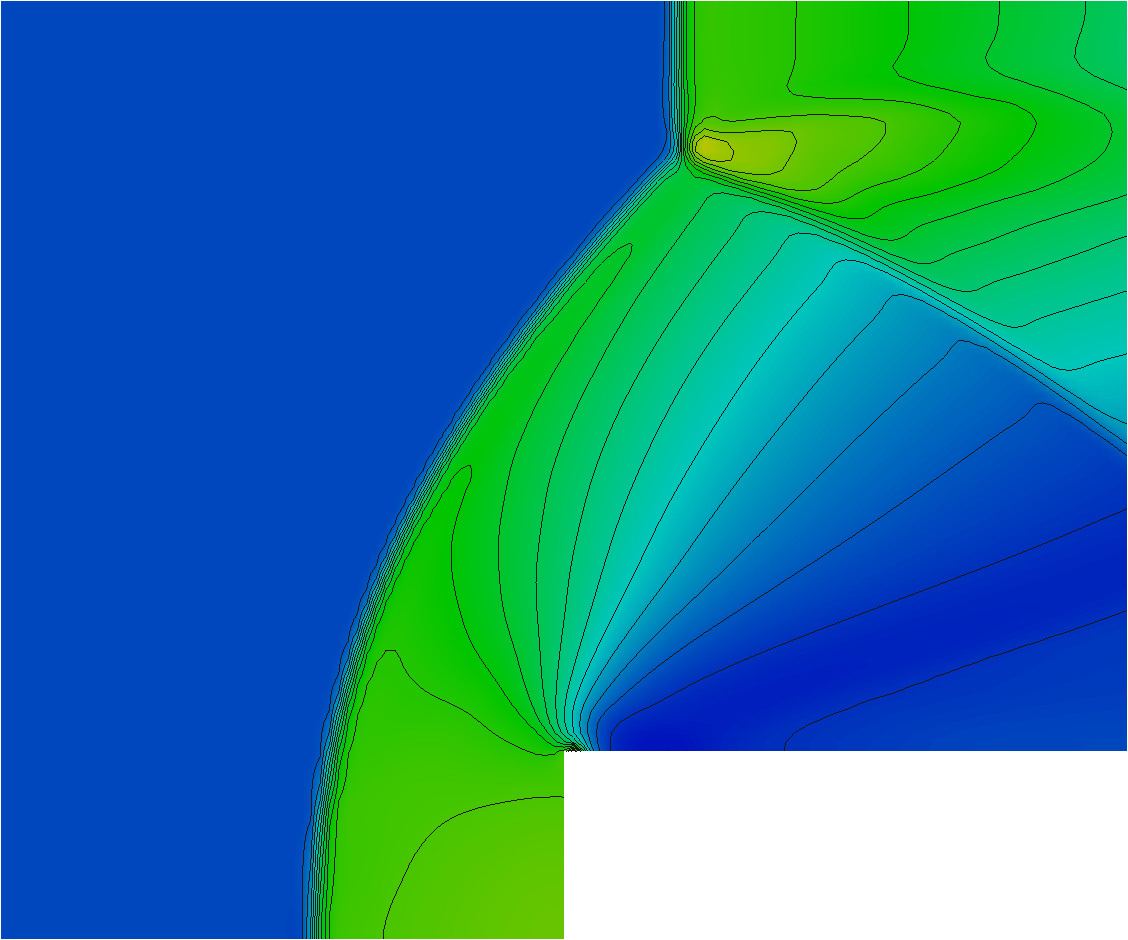}
    \caption{MST, GPS = $1/100$ (Mach reflection)}
    \label{fig:MST1By100}
  \end{subfigure}
  \begin{subfigure}{.6\textwidth}
    \centering
    \includegraphics[width=.9\textwidth]{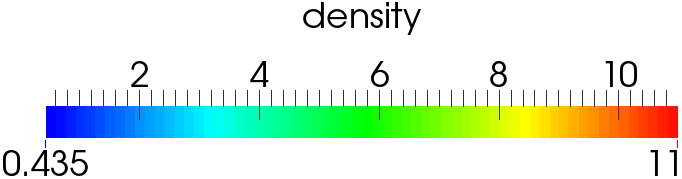}
    \caption{colour map}
    \label{fig:M40DX1By50MSTColMap}
  \end{subfigure}
\caption{Color plot of density with 15 equally spaced density contours for Mach 4.0 flow over a forward facing step, obtained using SSTCVD, MST and meshes with GPS of $1/50$, $1/100$. Using MST leads to Mach reflection at the
top wall.}
\label{fig:M40D10DX1By50MST}
\end{center}
\end{figure}

Figure~\ref{fig:M40D10DX1By800} has colour plots of density obtained using meshes with GPS of $1/800$ and all the five WBCTs mentioned in the previous section. All
the five
techniques result in a Mach reflection on the top wall (GE, see figure~\ref{fig:fowFacStepGeo}). Similar solutions are obtained for GPS of $1/400$.
Figure~\ref{fig:M40D10DX1By200} has the colour plots of density obtained using meshes with GPS of $1/200$. Apart from MST and SSTCVD, the other techniques fail to produce a Mach
reflection on the top wall, as they did for meshes with GPS of $1/400$ and $1/800$. Using MST resulted in a Mach reflection on the top wall for all tried GPS of $1/50, 1/100,
1/200, 1/400 ,\text{ and } 1/800$, whereas using SSTCVD did not produce a Mach reflection for GPS of $1/50$ and $1/100$, as shown in figure~\ref{fig:M40D10DX1By50MST}.

\begin{table}[!htbp]
  \begin{center}
    \caption{Shock standoff distances for Mach 4.0 flow over a forward facing step for different WBCTs. SSTNGW predicts the grid independent shock standoff distance even
      for coarse meshes. `-' indicates that 
  computations for using a mesh with this GPS were not done (as grid independence was already achieved).}
\label{tab:shkDstM40Cmp}
%      \begin{tabular}{ | p{0.15\textwidth} | p{0.08\textwidth} | p{0.08\textwidth} | p{0.08\textwidth} | p{0.08\textwidth} | p{0.08\textwidth} | }
      \begin{tabular}{ |>{\centering\arraybackslash}m{0.1\textwidth} | >{\centering\arraybackslash}m{0.08\textwidth} | >{\centering\arraybackslash}m{0.08\textwidth} | >{\centering\arraybackslash}m{0.08\textwidth} | >{\centering\arraybackslash}m{0.08\textwidth} | >{\centering\arraybackslash}m{0.08\textwidth} | >{\centering\arraybackslash}m{0.08\textwidth} |}
      \hline
      \multirow{2}{*}{WBCT} & \multicolumn{6}{c|}{Grid point spacing}\\\cline{2-7}
                                           &$1/50$ & $1/100$ & $1/200$ & $1/400$ & $1/800$ & $1/1600$ \\ 
      \hline
      SSTNGW & 0.246 & 0.246 & 0.246 & 0.246 & 0.246 & - \\ \hline
      SSTGW & 0.244 & 0.245 & 0.246 & 0.246 & 0.246 & - \\ \hline
      SSTNPE & 0.246 & 0.247 & 0.246 & 0.246 & 0.246 & -\\ \hline
      SSTCVD & 0.256 & 0.251 & 0.248 & 0.247 & 0.247 & - \\ \hline
      MST & 0.281 & 0.264 & 0.255 & 0.251 & 0.249 & 0.248 \\ \hline
    \end{tabular}
  \end{center}
\end{table}
Table~\ref{tab:shkDstM40Cmp} has the shock standoff distances (SB in figure~\ref{fig:fowFacStepGeo}) for the 5 WBCTs, for different grid point
spacings. The portion of shock near the wall AB is a normal shock with pre and post shock densities of $1.400$ units and $6.474$ units, respectively. The point S 
(shock location, in figure~\ref{fig:fowFacStepGeo}) is taken to be located on AB where the density is equal to half of the pre and post shock densities (which is equal
to $3.937$).
For SSTNGW, SSTGW, SSTNPE and SSTCVD, the shock standoff distance is grid independent for GPS of $1/400$ and $1/800$. The same is true for MST for GPS of $1/800$ and $1/1600$.
With this, the grid independence is achieved for all the WBCTs used.

It is evident from table~\ref{tab:shkDstM40Cmp} that SSTNGW predicts the grid independent shock standoff distance even for coarser meshes (with GPS of $1/50$). However, as
shown in figures~\ref{fig:M40D10DX1By800} and \ref{fig:M40D10DX1By200}, except for MST and SSTCVD, none of the other WBCTs are able to 
accurately capture the grid independent shock structure - that
of a Mach reflection on GE (see figure~\ref{fig:fowFacStepGeo}).
MST captures this shock structure even for GPS of $1/50$ and $1/100$, whereas SSTCVD does not, as shown in figure~\ref{fig:M40D10DX1By50MST}.

%1/200 MST machstem height =  (4.0 - 3.43)/4 = 0.1425
%1/200 SSTCVD machstem height =  (4.0 - 3.55)/4 = 0.1125
%1/800 MST machstem height =  (16.0 - 13.85)/16.0 = 0.134378
%1/800 SSTCVD machstem height =  (16.0-13.97)/16.0 = 0.126875
%1/800 noPenEnf machstem height =  (16.0-13.99)/16.0 = 0.125625
%1/800 noPenEnf machstem height =  (16.0-13.99)/16.0 = 0.125625
%1/800 symmConds machstem height =  (16.0-13.99)/16.0 = 0.125625
%1/800 noCornPt machstem height =  (16.0-13.99)/16.0 = 0.125625

%For coarser meshes, SSTNGW gives a better estimate of the shock standoff distance than the other corner techniques whereas MST gives a better prediction of the shock
%structure, namely an oblique
%shock with Mach reflection at the top wall GE(see Figure \ref{fig:fowFacStepGeo}).

Next, the WBCTs are used for computing flow over backward facing step.

\subsection{Supersonic flow over a backward facing step}
\begin{figure}[!htbp]
\begin{center}
  \includegraphics[width=.5\textwidth]{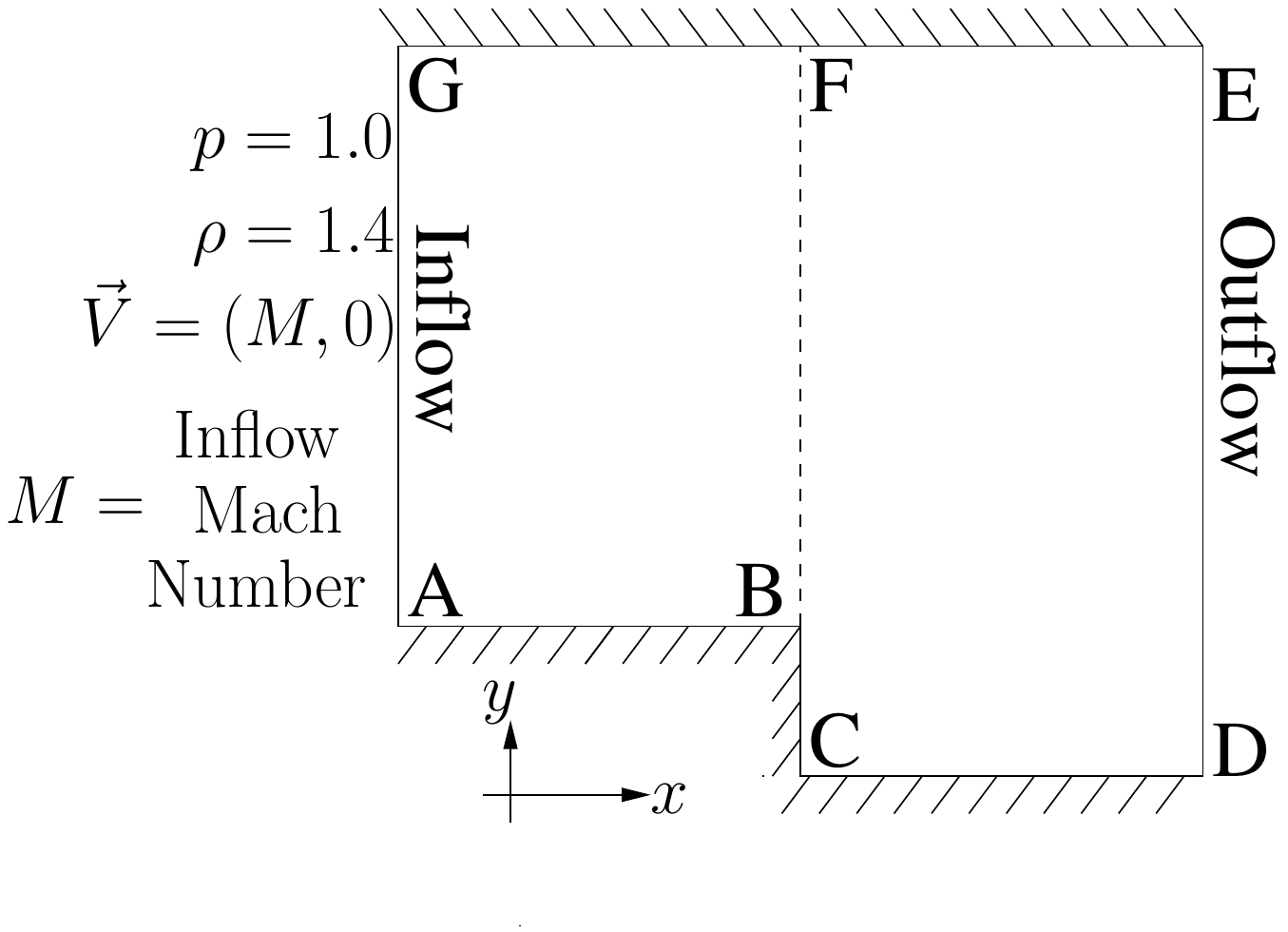}
  \caption{Backward facing step: Sketch of problem domain and boundary conditions (not drawn to scale).}
  \label{fig:BacFacStepGeo}
\end{center}
\end{figure}
Figure~\ref{fig:BacFacStepGeo} shows the sketch of problem domain and boundary conditions. 
The lengths of different portions of the flow field are (refer to figure~\ref{fig:BacFacStepGeo} for labels) - AB = $0.2$ units,
AG = $0.8$ units, BC = $0.2$ units, ED = $1.0$ units. 
Except for the inflow and outflow, all of the boundaries are free slip walls. The inflow is supersonic. For ease of applying boundary conditions, the length of CD is chosen so
that the outflow is also supersonic. In our computations, for different WBCTs, different values for CD in the range of $3.1$ units to
$4.8$ units were chosen. At the inflow boundary, the state is prescribed. At the outflow boundary one sided differences, biased in the negative $x$ direction
are used to calculate $x$ derivatives.

Conditions at inflow are: $\rho = 1.4$, $(u, v) = (M, 0)$, where $M$ is the inflow Mach number, and $p = 1.0$. Computations for two inflow conditions with Mach numbers of 1.5
and 2.5 were done using meshes with GPS of $1/50, 1/100, 1/200,\text{ and } 1/400$ for the five different WBCTs.

For both inflow Mach numbers of $1.5$ and $2.5$, the five WBCTs produce similar numerical solutions all having an expansion fan and reattachment shock, which
reflects off the top wall.
The solutions for inflow Mach number $1.5$
%, as can be seen in the color plots (\ref{fig:M15D10W3Pres}), visibly
differ in the position of the reattachment shock for different WBCT. The same trend appears for the inflow Mach number of 2.5.

Next, the flow leak due to using SST for this problem is described and it is compared with that of the Mach 4.0 flow over a forward facing step.

\subsubsection{Flow Leak}
\begin{figure}[!htbp]
  \begin{center}
    \includegraphics[width=0.8\textwidth]{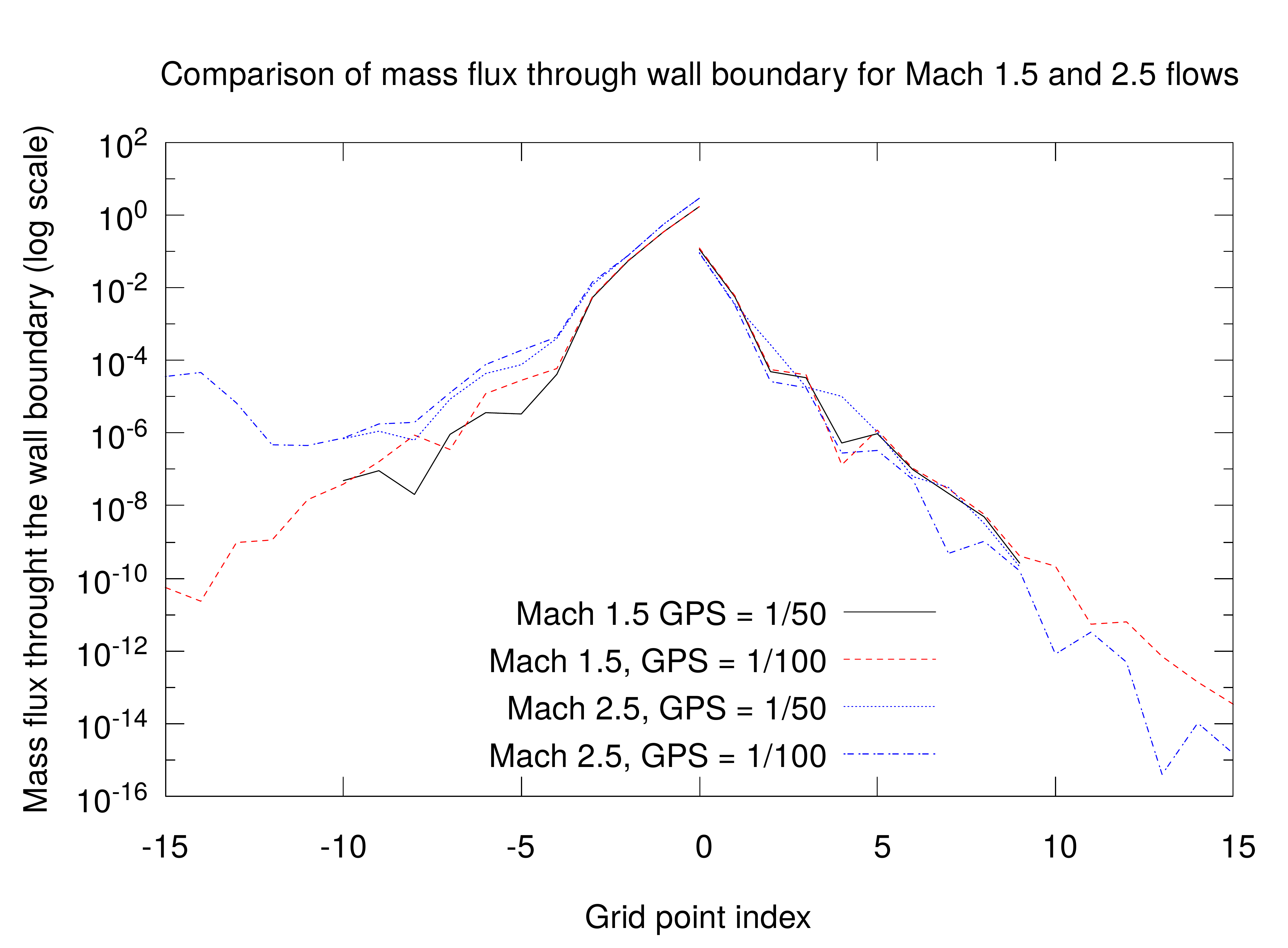}
    \caption{Plot of non-zero mass flux normal to the wall boundary, due to using SST near corner vs grid point index (indexing as shown in 
      figure~\ref{fig:gridPointAtCornerLab}), for Mach 1.5 and 2.5 flows over backward facing step,
    for grid point spacings of $1/50$, $1/100$.}
\label{fig:massLeakComparBacFacStepM15M25}
\end{center}
\end{figure}

Table~\ref{tab:M15massEnerLeakPercentage} has the mass and energy leak near the corner as a percentage of the inflow mass and energy, for the Mach 1.5 flow over a
backward facing step
for grid point spacings of $1/50$, $1/100$ and $1/200$. These are calculated similar to the data in table~\ref{tab:M40massEnerLeakPercentage} using trapezoidal rule.
The leak percentages are similar for the Mach 2.5 flow also. Figure~\ref{fig:massLeakComparBacFacStepM15M25} has plots
of mass flux on the wall boundary near the corner for the Mach 1.5 and 2.5 flows for comparison. 
\begin{table}[!htbp]
  \centering
  \caption{Mach 1.5 flow over backward facing step, mass and energy leak per unit time as a percentage of inflow mass and energy per unit time, respectively (Positive value indicates that mass or energy is flowing in). }
\label{tab:M15massEnerLeakPercentage}
%    \begin{tabular}{|p{0.05\textwidth}|p{0.14\textwidth}|p{0.145\textwidth}|p{0.075\textwidth}|p{0.14\textwidth}|p{0.145\textwidth}|p{0.075\textwidth}|}
  \begin{tabular}{|>{\centering\arraybackslash}m{0.05\textwidth}|>{\centering\arraybackslash}m{0.13\textwidth}|>{\centering\arraybackslash}m{0.145\textwidth}|>{\centering\arraybackslash}m{0.09\textwidth}|>{\centering\arraybackslash}m{0.13\textwidth}|>{\centering\arraybackslash}m{0.145\textwidth}|>{\centering\arraybackslash}m{0.09\textwidth}|}
      \hline
      \multirow{2}{5em}{GPS}&\multicolumn{3}{c|}{Mass leak rate percentage ($\dot{m}_l/\dot{m}_i\times 100$)}&\multicolumn{3}{c|}{Energy leak rate percentage ($\dot{e}_l/\dot{e}_i\times 100$)}\\\cline{2-7}
                            &Below corner ($a$)&Upstream of corner ($b$) $\times 10^{3}$& Total ($|a|+|b|$)&Below corner ($c$)&Upstream of corner ($d$) $\times 10^{3}$&Total ($|c|+|d|$)\\ 
      \hline
      $1/50$ & 0.27 & -3.3 & 0.28 & 0.37 & -4.9 & 0.38\\ \hline
      $1/100$ & 0.14 & -1.8 & 0.14 & 0.19 & -2.6 & 0.19\\ \hline
      $1/200$ & 0.07 & -0.8 & 0.07 & 0.09 & -1.3 & 0.09\\ \hline
    \end{tabular}
\end{table}

The mass and energy leak percentages for flow over backward facing step are similar to that of the Mach 4.0 flow over forward facing step, as can be seen from
the data in tables~\ref{tab:M15massEnerLeakPercentage} and \ref{tab:M40massEnerLeakPercentage}. A major portion of the leak happens below the corner for the flow over a
backward facing step, whereas for flow over forward facing step it happens downstream of the corner. For Mach 4.0 flow over forward facing step, the leak below the corner
is approximately one order of magnitude less than the leak downstream of the corner (see table~\ref{tab:M40massEnerLeakPercentage}), whereas for Mach 1.5 flow over
backward facing step, the leak upstream of the corner is approximately two orders of magnitude less than that below the corner.

\section{Conclusions}
The problems of state at the expansion corner point and flow leak due to using standard symmetry technique near corners were addressed. A method to tackle the
`corner point state' problem was proposed. Using MGW, it was shown that using SST (SSTGW) will lead to leak near the expansion corner (see figures~\ref{fig:massLeakComparM30}
and \ref{fig:enerLeakCompar})
and that refining the mesh will lead to reduction of flow leak near the corner as evidenced by the data in table~\ref{tab:M40massEnerLeakPercentage}.

To reduce the flow leak and to limit it to the corner point, three WBCTs - SSTNPE, SSTCVD and  modifications to the standard symmetry 
technique (MST) - were proposed  and implemented. The problem of leak at the corner still exists in the new WBCTs proposed. It is not clear how it can be
eliminated because the normal and tangent at the corner are not defined and simultaneous application of free slip and no-penetration at the corner is not possible.

Results obtained using the five different WBCTs for flows over forward facing and backward facing step were presented and compared.
Of the five WBCTs, for SSTNPE, SSTCVD and MST there is no mass leak at any grid point on the wall except the one at the expansion corner. Of SSTNPE, SSTCVD, and MST, only MST takes into
account the term
$\partial (\rho V_n V_{\tau})/\partial \tau$ (in equation~(\ref{eq:normMomentumEquation})) for enforcing no-penetration condition, while SSTNPE and SSTCVD do not.

%The resistance due to the wall increases going from SSTNGW to MST, which is evident from the shock standoff distances for Mach 3.0 flow over a step. For the Mach 3.0 flow over
%forward facing step, using SSTGW -SSTCVD and finer meshes led to the shock oscillating at the inflow, whereas using MST led to the shock oscillating near the inflow for all three
%meshes used.
%For the Mach 4.0 flow over forward facing step, the shock was in between the inflow and step for all meshes and corner techniques used.

For the Mach 4.0 flow over forward facing step, SSTNGW predicts the grid independent shock standoff distance for the coarser meshes also.
MST gives a better prediction of the shock structure (the type of shock reflection that happens at the wall GE, see figure~\ref{fig:fowFacStepGeo}).
Using MST, a Mach
reflection at the wall GE (see figure~\ref{fig:fowFacStepGeo}) and a $\lambda$ shock  was obtained for all grid
point spacings used (see figure~\ref{fig:M40D10DX1By50MST}). Whereas for the other corner techniques, only the finer meshes gave a solution with Mach reflection and
$\lambda$ shock (see figure~\ref{fig:M40D10DX1By800}). The solutions obtained with coarser meshes have regular shock reflection (see figure~\ref{fig:M40D10DX1By200}, 
\ref{fig:M40D10DX1By50MST}).

For the problem of flow over a backward facing step, the total mass and energy leak percentages due to using SST were similar to that for flow over
forward facing step. For flow over backward facing step the major portion of the leak happens below the corner whereas this happens downstream of the corner for flow over
forward facing step.

\bibliography{research}
\bibliographystyle{elsarticle/elsarticle-harv}

\end{document}